\documentclass[manuscript]{copernicus}
\nolinenumbers

\usepackage{hyperref}

\usepackage{natbib}

\usepackage{graphicx}
\usepackage{wrapfig}

\usepackage{float}

\usepackage{mathtools}
\usepackage{amsmath}
\usepackage{amsfonts}

\usepackage{makecell}
\usepackage{multirow}
\usepackage{booktabs}

\usepackage{array}
\newcolumntype{P}[1]{>{\centering\arraybackslash}m{#1\textwidth}}

\usepackage[dvipsnames, table]{xcolor}
\newcommand{\teal}[1]{{#1}}
\newcommand{\blue}[1]{{#1}}

\definecolor{lightgrey}{RGB}{240,240,240}
\definecolor{lightblue}{RGB}{200,230,250}

\usepackage{amsthm}
\usepackage{enumitem}
\usepackage{placeins}
\usepackage{overpic}
\usepackage[export]{adjustbox}
\usepackage{mathrsfs}

\begin{document}

\title{The stochastic skeleton model for the Madden-Julian Oscillation with time-dependent observation-based forcing}


\Author[1,2]{No\'{e}mie}{Ehstand} 
\Author[3,4]{Reik V.}{Donner}
\Author[1]{Crist\'{o}bal}{L\'{o}pez}
\Author[5]{Marcelo}{Barreiro}
\Author[1][emilio@ifisc.uib-csic.es]{Emilio}{Hern\'{a}ndez-Garc\'{\i}a}

\affil[1]{Instituto de F\'{\i}sica Interdisciplinar y Sistemas
Complejos (IFISC), CSIC-UIB, Campus Universitat de les Illes
Balears, E-07122 Palma de Mallorca, Spain} \affil[2]{Present
address: Institut Terre et Environnement de Strasbourg (ITES),
Universit\'{e} de Strasbourg, 5 rue Descartes, Strasbourg
F-67084, France} \affil[3]{Department of Water, Environment,
Construction and Safety, Magdeburg-Stendal University of
Applied Sciences, Breitscheidstra{\ss}e 2, D-39114 Magdeburg,
Germany} \affil[4]{Research Department IV-Complexity Science,
Potsdam Institute for Climate Impact Research (PIK) - Member of
the Leibniz Association, Telegrafenberg A31, D-14473 Potsdam,
Germany} \affil[5]{Departamento de Ciencias de la Atm\'{o}sfera
y F\'{\i}sica de los Oceanos, Instituto de F\'{\i}sica,
Facultad de Ciencias, Universidad de la Rep\'{u}blica, Igua
4225, 11400 Montevideo, Uruguay}



\runningtitle{TEXT}

\runningauthor{TEXT}

\received{}
\pubdiscuss{} 
\revised{}
\accepted{}
\published{}


\firstpage{1}

\maketitle

\begin{abstract}
    We analyze solutions to the stochastic skeleton model, a minimal nonlinear oscillator model for the
    Madden-Julian Oscillation (MJO).  This model has been recognized for its ability to reproduce several
    large-scale features of the MJO. In previous studies, the model's forcings were predominantly chosen
    to be mathematically simple and time-independent. Here, we present solutions to the model with
    time-dependent observation-based forcing functions. Our results show that the model, with these
    more realistic forcing functions, successfully replicates key characteristics of MJO events,
     such as their lifetime, extent, and amplitude, whose statistics agree well with observations.
     However, we find that the seasonality of MJO events and the spatial variations in the MJO
     properties are not well reproduced. {Having implemented the model in the presence of time-dependent
     forcings, we can analyze the impact of temporal variability at different time scales. In particular,
     we study the model's ability} to reflect changes in MJO characteristics under the different phases
     of ENSO. We find that it does not capture significant differences in the studied characteristics of
     MJO events in response to differences in conditions during El Ni\~no, La Ni\~na, and neutral ENSO.
\end{abstract}


\section{Introduction}
The Madden--Julian Oscillation \citep{Madden1972} is the dominant component of intra-seasonal variability in the tropics \citep{Woolnough2019}. It is a planetary-scale wave envelope of smaller-scale convective processes, slowly propagating eastward along the equator, with a period of 40 to 50 days on average, and a mean propagation speed of about $5$ m/s, with fluctuations ranging from $1$ to $9$ m/s \citep{Chen2020}. As it propagates, the MJO causes disturbances in rainfall and winds which impact weather and climate in both the tropics and the extratropics \citep{Zhang2013}. In particular, the MJO plays a significant role in the occurrence of various weather extremes such as tropical cyclones, tornadoes, extreme rainfall events and extreme surface temperatures \citep{Jones2004,Pohl2006,Julia2012,Thompson2013,Zhang2013,Jeong2005}.\bigbreak

With its slow eastward propagation on intraseasonal time scales, the MJO is a key source of subseasonal predictability \citep{Woolnough2019,Vitart2019} and is of utter importance to enhance preparedness for such extreme events. Accurately forecasting the MJO has thus become a foremost goal in subseasonal-to-seasonal forecasting. But, even with significant progress in recent years \citep{Lim2018,Kim2019, Silini2021}, its representation and predictability in numerical models continues to be a challenging task \citep{Liu2024}. Moreover, a recent study suggests that there is an increased probability of extinction of the MJO after 27 days, which seems to point to an internal mechanism of exhaustion rather than to the effect of an external barrier \citep{Corral2023}. \bigbreak

The underlying physical mechanisms that govern the MJO are not yet fully understood \citep{Zhang2013}. A deeper understanding of these mechanisms is however essential for improving MJO predictions, as they form the foundation for the development of more accurate models. This knowledge gap has motivated the development of data-driven models \citep{Diaz2023} and simplified low-dimensional physics-based models \citep{Zhang2020} that aim to capture the key aspects of the MJO's behavior. One such model, known as the skeleton model, provides a framework for investigating some essential dynamics of the MJO. \bigbreak

The skeleton model is a minimal nonlinear oscillator model for the Madden-Julian Oscillation, introduced by \cite{Majda2009b}. The model is derived from the well-known primitive equations for the dry atmosphere in the tropics complemented with a modelling of moist convection. It is an idealized model with a minimal number of parameters and a single quadratic nonlinearity. Nonetheless, it provides insights into atmospheric dynamics when coupled with moist convection in the tropics and is able to reproduce large-scale features of the MJO, including its slow eastward propagation at about $5$m/s, its near-zero group velocity and its horizontal quadrupole vortex structure. Several versions of the model have been developed based on the initial work from Majda and Stechmann \citep{Majda2011,Thual2014,Thual2015a,Thual2015}. In particular, adding stochasticity to the initial model, \cite{Thual2014} showed that the skeleton theory also captures the intermittent generation of MJO events and their organization into wave trains with growth and decay. \bigbreak

The model includes two forcing functions, representing latent heating and radiative cooling in the tropics. In the majority of previous studies, these functions were chosen to be mathematically simple, time-independent and identical \citep{Majda2009b,Majda2011,Thual2014,Thual2015,Stachnik2015,Chen2016}. The aim of the present study is to assess whether the model forced with more dynamic and observationally grounded functions can still accurately reproduce key characteristics of MJO events. By using more realistic forcings, our work offers a more robust test of the skeleton model's applicability and relevance in understanding the MJO dynamics as, simplified, static forcings might limit the realism of the model's outputs. \blue{Our goal is not to predict individual MJO events, which is even a challenge for models of much higher complexity \citep{Jiang2020, Zhou2024} but to check how well the skeleton model reproduces their statistical properties.} \bigbreak

Precisely, we consider the stochastic skeleton model as presented in \cite{Thual2014} with forcing functions computed from observational and reanalysis data following a methodology presented in \cite{Ogrosky2015}. Further, while \cite{Ogrosky2015} forced the model with long-term averages of the computed functions, in this work, we keep their time-dependence. Thus, we present solutions to the model when the latent heating and radiative cooling functions are observation-based, time-dependent, non-identical functions. To our knowledge this is the first study of the MJO skeleton model with forcing functions combining these three characteristics.

\blue{Further, several studies suggested that the sea surface temperature (SST) variability at interannual and longer time scales influences the MJO's characteristics. In particular the variability associated with the El Ni\~no--Southern Oscillation (ENSO) modulates the extent of MJO's eastward propagation \citep{Kessler2001,Tam2005,Pohl2007}, its lifetime \citep{Pohl2007} and its speed \citep{Wei2019,Diaz2023}.} With our implementation of the skeleton model using time-dependent forcings, we can now explore the ability of the MJO skeleton model to induce differences in selected MJO characteristics under El Ni\~no, La Ni\~na and neutral ENSO conditions.
\bigbreak

This paper is organized as follows. In Section
\ref{sec:skeleton_model_presentation} we present the model and
in Section \ref{sec:forcing-profiles} the computation of the
forcing functions from observational and reanalysis data is
explained. The identification of MJO events in the model is
performed based on an objective index: the skeleton
multivariate MJO (SMM) index. The computation of the SMM index
is described in Section \ref{sec:rmm_index}. The numerical
solutions to the model are then presented in Section
\ref{sec:skel_model_num_sol}. The identification of MJO events
is briefly illustrated in Section \ref{sec:ID_MJO}. Then, in
Section \ref{sec:obs_sim}, we compare statistics of MJO
characteristics between observations and simulations. Finally,
in Section \ref{sec:ENSO_mjo_influence}, we present the
statistics of selected characteristics of MJO events in
observations and in the stochastic skeleton model simulations
under El Ni\~no, La Ni\~na and neutral ENSO conditions.

\section{The MJO skeleton model}\label{sec:skeleton_model_presentation}
\subsection{Deterministic model}

The MJO skeleton model \citep{Majda2009b,Majda2011} combines the linear, long-wave scaled, primitive equations \citep[see][]{White2003,Vallis2017} with a conservation equation for the moisture and a dynamic equation describing the interactions between the lower tropospheric moisture anomaly and the planetary-scale envelope of convective activity:
\begin{equation}
\begin{split}
    u_t -yv &= -p_x,\\ 
    yu &= -p_y,\\ 
    0 &= -p_z + \theta,\\ 
    u_x + v_y + w_z &= 0,\\ 
    \theta_t + w &=\bar{H}a - s^{\theta},\\ 
    q_t - \tilde{Q}w &= -\bar{H}a+s^{q},\\ 
    a_t &= \Gamma qa,
\end{split}
\label{eq:Skeleton-model-complete}
\end{equation}

\noindent where $x$, $y$, and $z$ are the zonal, meridional and vertical coordinates, and $u$, $v$ and $w$ are the velocity anomalies in these directions, respectively, $p$ and $\theta$ are the pressure and potential temperature anomalies, $q$ is the lower tropospheric moisture anomaly, and $a$ is the envelope of convective activity. Note that all variables are anomalies {from a radiative-convective equilibrium} except for $a$. The variables $s^{\theta}$ and $s^{q}$ represent external sinks/sources of temperature and moisture, such as radiative cooling and latent heating, respectively. They act as forcing in the model and will be described in more detail in Section \ref{sec:forcing-profiles}. The model has only three parameters: $\tilde{Q}$ is the mean background vertical moisture gradient, $\Gamma$ represents the sensitivity of convective activity tendency to moisture anomalies and $\bar{H}$ is a scaling constant for the convective activity. The equations have been non-dimensionalized using some standard equatorial length and time scales \citep{Majda2009a}. The first five equations of the model describe the dry dynamics of the atmosphere (the conservation of horizontal momentum \blue{in $x$ and $y$}, the hydrostatic balance, the conservation of mass and the conservation of potential temperature). The sixth equation describes the conservation of low level moisture. The last equation is the non-linear interaction between moisture and convection. It entails the idea that the moisture anomalies influence the growth and decay rates of the planetary-scale envelope of convective activity.

To obtain the model in its simplest version \blue{\citep[Section~2.3.3]{Majda2009b,Majda2011,Thual2014, Majda2019}}, Equations \eqref{eq:Skeleton-model-complete} are truncated in the vertical and meridional directions. In the vertical, the variables are expanded in terms of sines and cosines, keeping only the first baroclinic mode, i.e. $u(x,y,z,t) \approx u_1(x,y,t)\sqrt{2}\cos(z)$, $v(x,y,z,t) \approx v_1(x,y,t)\sqrt{2}\cos(z)$, $p(x,y,z,t) \approx p_1(x,y,t)\sqrt{2}\cos(z)$, $w(x,y,z,t) \approx w_1(x,y,t)\sqrt{2}\sin(z)$, $\theta(x,y,z,t) \approx \theta_1(x,y,t)\sqrt{2}\sin(z)$ \citep[see][]{Khouider2012}. Here, {$z\in [0,\pi]$ in non-dimensional units, and $z\in[0,H_{top}]$ in dimensional units, where $H_{top}$ is the height of the tropopause.} Further, it is assumed that $q=q_1(x,y,t)\sqrt{2}\sin(z)$, $a=a_1(x,y,t)\sqrt{2}\sin(z)$, $s^\theta=s^\theta_1(x,y,t)\sqrt{2}\sin(z)$ and $s^q=s^q_1(x,y,t)\sqrt{2}\sin(z)$ \citep[see][]{Majda2015}. Dropping the subscript $1$ for simplicity (i.e. $u_1 \rightarrow u$, $v_1 \rightarrow v$, etc.), System \eqref{eq:Skeleton-model-complete} becomes:
\begin{equation}
\begin{split}
    u_t -yv -\theta_x &= 0,\\ 
    yu - \theta_y &= 0, \\ 
    \theta_t - u_x - v_y &=\bar{H}a - s^{\theta},\\ 
    q_t + \tilde{Q}(u_x + v_y) &= -\bar{H}a+s^{q},\\ 
    a_t &= \Gamma qa.
\end{split}
\label{eq:Skeleton-model-complete-simplified}
\end{equation}

In the meridional direction, the variables and forcing functions are expanded using parabolic cylinder functions $\{\phi_m(y)\}$ such that, $u(x,y,t) = \sum_m u_m(x,t)\phi_m(y) $, etc., where the first three modes have the form
$ \phi_0(y) = \pi^{-1/4}\exp(-y^2/2)$, $\phi_1(y) = \pi^{-1/4}\sqrt{2}y \exp(-y^2/2)$, $\phi_2(y) = \pi^{-1/4}(1/\sqrt{2})(2y^2-1)\exp(-y^2/2)$.
This expansion facilitates a change of variable in the dry dynamics (rows 1-4 of Eq. \eqref{eq:Skeleton-model-complete-simplified}) which allows to introduce new variables representing equatorial waves. In the simplest version of the model, only the amplitudes of the first mode of equatorial Kelvin wave structure ($K$) and equatorial Rossby wave structure ($R$) are kept, defined as
$K\equiv\frac{u_0-\theta_0}{\sqrt{2}}$ and $R\equiv u_2 - \theta_2 - \frac{u_0+\theta_0}{\sqrt{2}}$. In addition, it is assumed that the envelope of convection/ wave activity $a$ takes the form $a(x,y,t)= A(x,t)\phi_0(y) = [A_s(x) + A^*(x,t)]\phi_0(y)$ with $A_s$ representing the background state and $A^*$ fluctuations around this state, and that $q(x,y,t)$, $s^{\theta}(x,y,t)$ and $s^q(x,y,t)$ are truncated at the first mode: $q(x,y,t)=Q(x,t)\phi_0(y)$, $s^\theta(x,y,t)=S^\theta(x,t)\phi_0(y)$ and $s^q(x,y,t)=S^q(x,t)\phi_0(y)$. The final truncated equations then read:
\begin{equation}
    \begin{split}
        K_t + K_x &= -\frac{1}{\sqrt{2}}(\bar{H}A-S^\theta),\\
        R_t - \frac{1}{3}R_x &= -\frac{2\sqrt{2}}{3}(\bar{H}A-S^\theta),\\
        Q_t + \frac{1}{\sqrt{2}}\Tilde{Q}K_x - \frac{1}{6\sqrt{2}} \Tilde{Q} R_x &= \frac{\Tilde{Q}}{6}(\bar{H}A - S^\theta) - (\bar{H}A - S^q),\\
        A_t &= \gamma\Gamma Q (A_s + A^*),
    \end{split}
    \label{eq:skeleton-model-numerics}
\end{equation}
where all the functions depend now only on $x$ and $t$, and $\gamma = \int (\phi_0)^3 dy\approx 0.6$ results from the meridional projection of the non-linear equation.

The variables $u$, $v$ and $\theta$ can be approximately recovered via
\begin{equation}
    \begin{split}
    u(x,y,t) &= \frac{1}{\sqrt{2}} \left[ K(x,t) - \frac{1}{2}R(x,t) \right]\phi_0(y) + \frac{1}{4}R(x,t)\phi_2(y),\\
    v(x,y,t) &= \left[ \frac{1}{3}\partial_xR(x,t) - \frac{1}{3\sqrt{2}}(\bar{H}A(x,t)-S^\theta(x,t)) \right]\phi_1(y),\\
    \theta(x,y,t) &= -\frac{1}{\sqrt{2}}\left[K(x,t) + \frac{1}{2}R(x,t)\right]\phi_0(y) - \frac{1}{4}R(x,t)\phi_2(y).
    \end{split}
    \label{eq:uvtheta}
\end{equation}

\subsection{Stochastic model}
\label{subsec:stoch-mod}

In the skeleton model, the MJO is initiated and sustained by the synoptic (sub-planetary) scale convective activity patterns, which are considered collectively via their planetary-scale envelope $a$. These synoptic scale processes include for instance deep convective clouds which are highly irregular, intermittent and with low predictability. To account for such processes, \cite{Thual2014} proposed a modified version of the skeleton model, where the last equation in \eqref{eq:Skeleton-model-complete} is replaced by a stochastic process. The authors showed that this \emph{stochastic skeleton model} is able to generate intermittent MJO wave trains with growth and decay as observed in reality. Specifically, the variable $a$ in System \eqref{eq:Skeleton-model-complete} \blue{is replaced at each point by an independent random variable} taking discrete values separated by $\Delta a$, that is $a = \eta \Delta a$, with $\eta \in \mathbb{N}$. The evolution of $a$ is controlled by a birth-death process allowing for intermittent transitions between states $\eta$. This process is described by the following master equation for the probability of $\eta$, $P(\eta,t)$:
\begin{equation}
    \partial_t P (\eta) = [\lambda(\eta-1)P(\eta-1) - \lambda(\eta)P(\eta)] + [\mu(\eta +1)P(\eta+1) - \mu(\eta)P(\eta)]
    \label{eq:mean-field-equation}
\end{equation}
\noindent where $\lambda$ is the upward rate of transition and $\mu$ the downward rate. The choice of $\lambda$ and $\mu$ is made such that the dynamics of the non-stochastic skeleton model is recovered on average \citep[see][]{Thual2014}.

\section{Observation-based time-dependent forcing functions}
\label{sec:forcing-profiles}

As mentioned in the introduction, the majority of previous
studies on the MJO skeleton model used idealized,
time-independent and equal forcing functions $s^\theta$
(radiative cooling) and $s^q$ (latent heating), i.e.
$s^\theta(\mathbf{x}) = s^q(\mathbf{x})$
\citep[see][]{Majda2009b,Majda2011,Thual2014,Thual2015,Stachnik2015,Chen2016}.
\cite{Thual2014b} first studied the solutions of the skeleton
model with periodic variations in the forcing. The authors used
an idealized warm pool state representation of
$s^\theta(\mathbf{x},t)=s^q(\mathbf{x},t)$ migrating seasonally
in the meridional direction. In \cite{Ogrosky2015} and later in
\cite{Ogrosky2017}, the authors computed the forcing functions
based on long-term means of observational and reanalysis data,
leading to more realistic functions where $s^\theta(\mathbf{x})
\neq s^q(\mathbf{x})$.

Here, we consider the stochastic skeleton model, with forcing functions computed from observational and reanalysis data following the methodology presented in \cite{Ogrosky2015}. However, unlike Ogrosky and Stechmann, we do not take long-term averages but consider monthly varying data. We are hence concerned with solutions of model \eqref{eq:Skeleton-model-complete} when $s^\theta=s^\theta(\mathbf{x},t)$, $s^q=s^q(\mathbf{x},t)$ and $s^\theta \neq s^q$, that is, when the forcings are realistic (observation-based), time-dependent and non-identical. \blue{We stress that all model parameters were chosen as in \cite{Ogrosky2015}. No further parameter tuning has been performed, to show more clearly the effect of the new ingredient introduced here: the time-dependent forcing.} This section explains the computation of the profiles.

\subsection{Data sources}

To estimate the forcing terms, we use NCEP/NCAR reanalysis
latent heat net flux \citep{Kalnay1996} for the computation of
the latent heating $s^q$ and NCEP Global Precipitation
Climatology Project (GPCP) data \citep{NCEPPrecip,Huffman2001}
for the computation of $\bar{H}a$ (which then enters in the
calculation of both $s^q$ and $s^\theta$ as explained below).
The chosen data sub-sets cover the period 1979-2021 with a
monthly resolution. Both fields have global spatial coverage,
with a resolution of $1.875^\circ \times 1.875^\circ$ (degrees
of latitude and longitude) for the latent heat flux data set
and $2.5^\circ \times 2.5^\circ$ for the precipitation data
set.

\subsection{Estimation procedure}

First, following \cite{Ogrosky2015}, the 2D field $\bar{H}a(x,y)$ is computed using the formula
\begin{equation}
\bar{H}a = \left( \frac{g \rho_w L_v}{p_0 c_p} \right)  M ,
\end{equation}
where $M$ [m] represents the monthly precipitation data, $g=9.8$ m/s$^2$ is the gravitational acceleration constant, $\rho_w=10^3$ kg/m$^3$ is the density of water, $L_v=2.5 \cdot 10^6$ J/kg is the latent heat of vaporization, $c_p=1006$ J/(kg$\cdot$K) is the specific heat of dry air at constant pressure, and $p_0 = 1.013 \cdot 10^{5}$ kg m$^{-1}$ s$^{-2}$ is the mean atmospheric pressure at mean sea level. The above formula describes the rate at which the temperature of a column of air increases from the energy released by precipitation at a given location.
To obtain the 1D equatorial profile needed for the truncated model (see Section \ref{sec:skeleton_model_presentation}, Eq. \eqref{eq:skeleton-model-numerics}), the 2D field is projected onto the leading meridional mode $\phi_0$:
\begin{equation}
\bar{H}A(x,t) = \int_{-\infty}^{\infty}\bar{H}a(x,y,t)\phi_0(y)dy.
\end{equation}

Second, still following \cite{Ogrosky2015}, we compute the 1D equatorial forcing profile $S^q(x,t)$. The latent heat flux (LHF) is projected onto $\phi_0$:
\begin{equation}
LHF_0(x,t) = \int_{-\infty}^{\infty}LHF(x,y,t)\phi_0(y)dy,
\end{equation}
and $S^q$ is computed according to \cite{Ogrosky2015} as
\begin{equation}
S^q = H_{LHF} \cdot LHF_0,
\label{eq:Sq}
\end{equation}
where $$H_{LHF} \approx \frac{\langle \bar{H}A \rangle_{t,x} }{\langle LHF_0 \rangle_{t,x} } \approx 0.0067 \text{ K day}^{-1}\text{(Wm}^{-2}\text{)}^{-1},$$
with $\langle \cdot \rangle_{t,x}$ representing the time and zonal mean, {implying that $\langle \bar{H}A \rangle$ balances $\langle S^q \rangle$.}

In fact, according to \cite{Ogrosky2015} (Eq. (7) in that paper), for a steady-state solution to exist in the skeleton model, which has no damping, we must have
\begin{equation}
\langle \bar{H}A_s \rangle_x = \langle S_s^q \rangle_x = \langle S_s^\theta \rangle_x,
\label{eq:mjo_sm_condtion1}
\end{equation}
where $\langle \cdot \rangle_x$ represents the zonal mean and the subscript $s$ indicates the background state of the quantities, defined as their long-time average. From Eq. \eqref{eq:Sq} we see that this is satisfied for $\bar{H}A$ and $S^q$.

In addition, as explained in \cite{Ogrosky2015} (Eq. (8) in that paper), the model background convective activity must satisfy
\begin{equation}
    \bar{H}A_s = \frac{S_s^q - \tilde{Q}S_s^\theta}{1-\tilde{Q}}.
    \label{eq:condition_back_conv_ac}
\end{equation}
 To make sure that this is the case, we compute $S^\theta(x,t)$ as:
\begin{equation}
    S^\theta =  \frac{1}{\tilde{Q}}S^q - \frac{(1-\tilde{Q})}{\tilde{Q}} \bar{H}A.
    \label{eq:Stheta_equ}
\end{equation}

Note that, computed in this way, $S^\theta$ also automatically satisfies condition \eqref{eq:mjo_sm_condtion1}.

The computed functions $A_s$, $S^q$ and $S^\theta$ can be decomposed into spatial Fourier modes. In order to focus on planetary-scale variations, only the first 8 Fourier modes are kept.\bigbreak

\blue{The aim of this work} is to study the solutions to the MJO skeleton model when the forcing functions are \emph{time-dependent} observation-based functions. Therefore, while \cite{Ogrosky2015} used long-term averages of the computed $S^q$ and $S^\theta$, here we skip this step and keep the time dependence of the profiles. \teal{As mentioned above, the data sets have monthly resolution. Nonetheless, it was observed that using monthly-varying profiles causes a drop in the frequency of model-generated MJO events. Hence, as the model seems to require a degree of persistence in duration and amplitude of the forcing to adequately generate MJO events, we smooth the forcing functions in time using a three-month running mean. Further clarifying the specific factors that cause the studied conceptual model's failure to exhibit a reasonable frequency of MJO events when employing the time-dependent forcing without smoothing, as well as the impact of the smoothing time scale on the statistical properties of the generated events, requires further numerical as well as analytical study of the model. We suggest that those points should be clarified in targeted follow-up studies.}

\begin{figure}
\centering
\includegraphics[width = 0.55\textwidth]{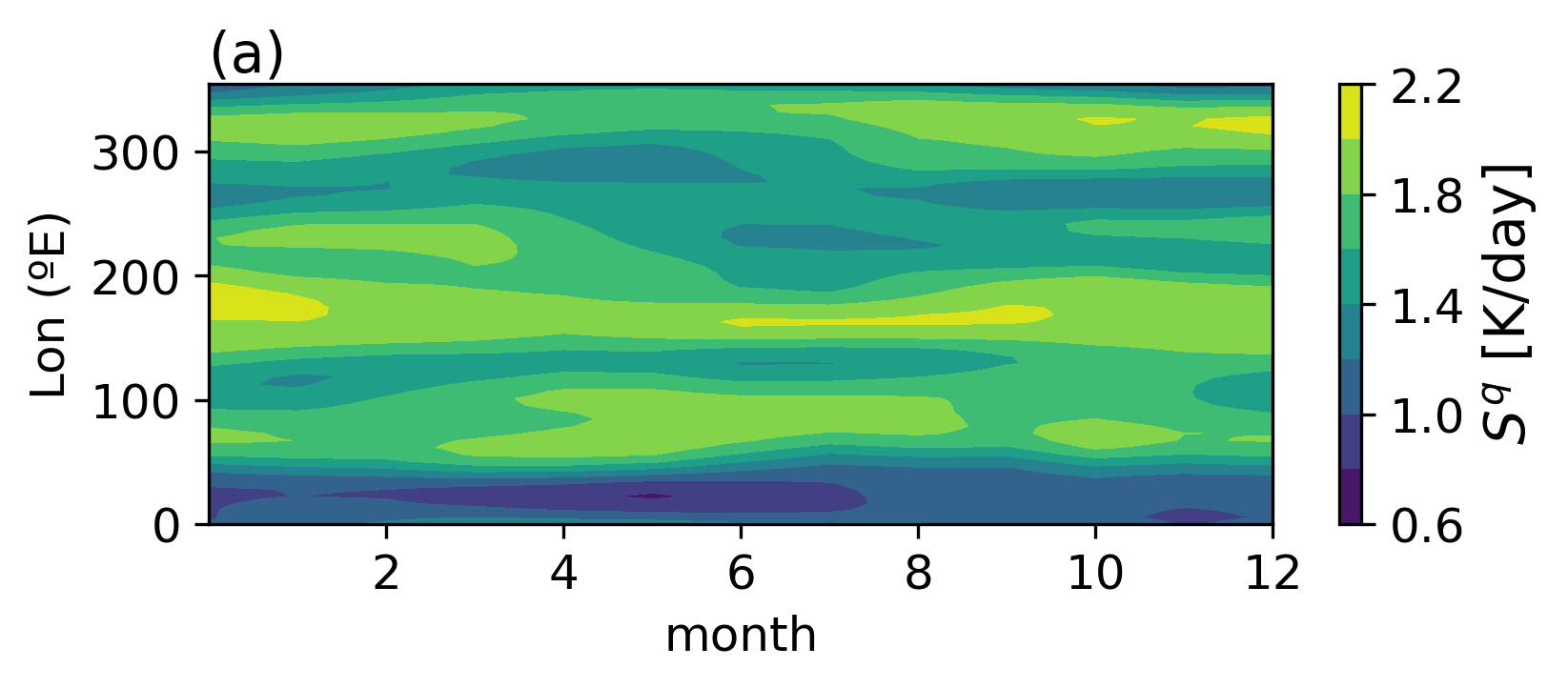}\\
\includegraphics[width = 0.55\textwidth]{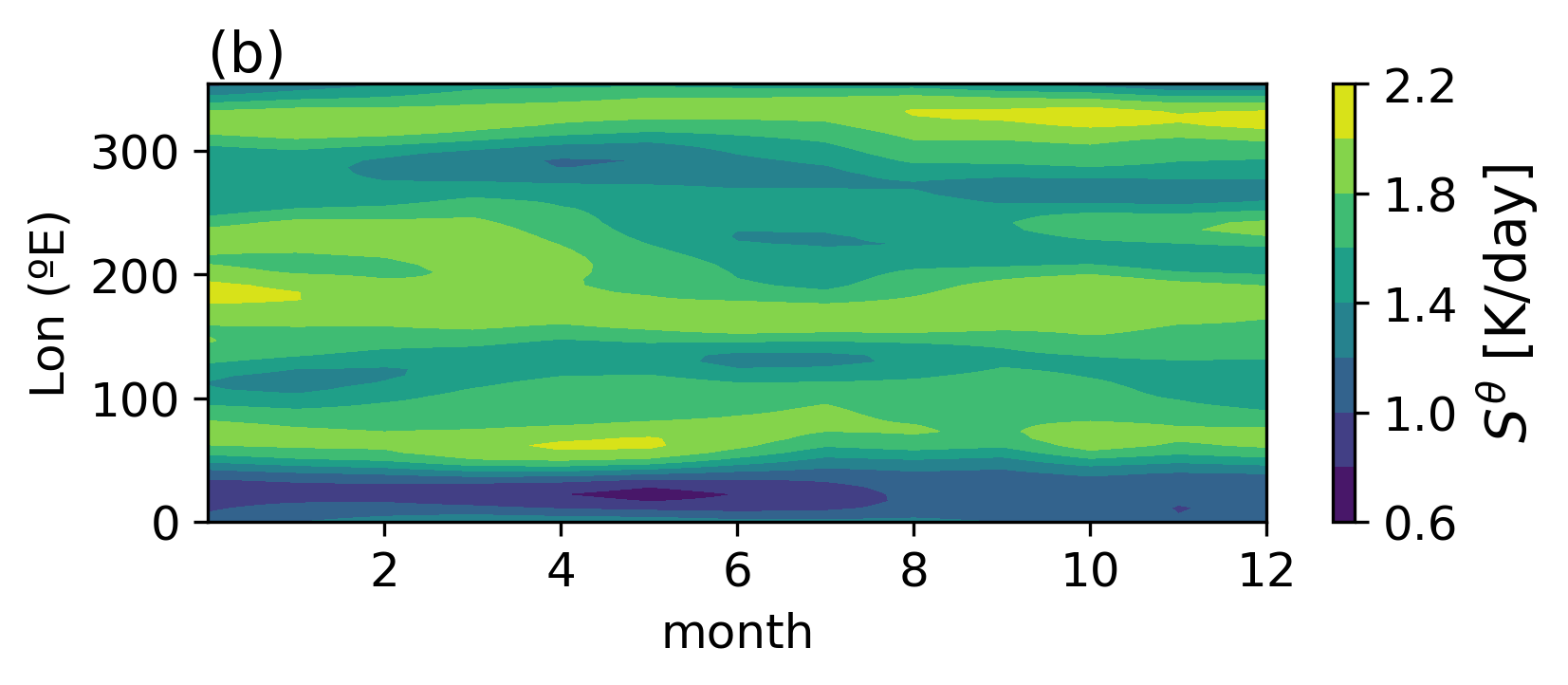}
\caption{Evolution of (a) the latent heating profile $S^q$ estimated from latent heat flux and (b) the radiative cooling profile $S^\theta$ computed according to Eq. \eqref{eq:Stheta_equ}. The abscissa indicates months of the year 1979.}
\label{fig:forcing-profiles}
\end{figure}

\teal{The three-month smoothed profiles are finally interpolated to the time step of the model.} We therefore obtain smooth time-dependent observation-based forcing functions $S^q$ and $S^\theta$. As an example, their evolution over the year 1979 is illustrated in Fig. \ref{fig:forcing-profiles}.


\section{Identification of MJO events: the skeleton multivariate MJO index}
\label{sec:rmm_index}

In observations, the most commonly used index to monitor the
MJO is the real-time multivariate MJO (RMM) index developed by
\cite{Wheeler2004}. It is based on an empirical orthogonal
function (EOF) analysis of the daily observed Outgoing Longwave
Radiation (OLR) as well as lower tropospheric (850 hPa) and
upper tropospheric (200 hPa) zonal winds, averaged meridionally
in the equatorial region between 15$^\circ$S-15$^\circ$N. {The
index is typically represented in a two-dimensional phase space
defined by the two dominant principal components (RMM1 and
RMM2). This space is divided into 8 sectors, each
corresponding to a distinct phase of the MJO cycle as its
convective center travels from the Indian Ocean across the
Pacific and towards the Western Hemisphere.}\bigbreak

In order to objectively identify MJO structures in the model
output, we compute an index similar to the RMM index following
the methodology presented in \cite{Stachnik2015}. The main
difference is that while RMM uses three variables (upper and
lower tropospheric winds and OLR), the model index is based on
a bivariate analysis with the (model) zonal wind $u$ as a
direct substitute for the lower tropospheric wind and the
negative convective heating $-\bar{H}a$ as a proxy for OLR.
Since clouds affect the radiation emitted at the top of the
atmosphere, OLR is often chosen as an indicator of cloudiness.
\cite{Wheeler2004} indeed used OLR as a proxy for convective
activity, justifying the choice of $-\bar{H}a$
herein.\footnote{\cite{Stechmann2014a} showed that OLR
variations are proportional to the total diabatic cooling
variations in the atmosphere. This might be approximated in the
model as the negative of the sum of latent (convective) heating
and radiative cooling  $-(\bar{H}a - s^\theta)$. Nonetheless,
we choose to use $-\bar{H}a$ alone as a proxy for OLR, since
the essential aim is to represent the equatorial convective
activity.}  The steps for the computation of this
\emph{skeleton multivariate MJO} (SMM) index are briefly
described below and full details can be found in
\cite{Stachnik2015}.

We first isolate the intraseasonal signal in the data. To do so
the daily anomalies of $u$ and $-\bar{H}a$ are filtered using a
$20-100$ days Lanczos filter. Second, each field is normalized
by its global standard deviation so that they have an equal
contribution in the computation of empirical orthogonal
functions. Lastly, as in \cite{Wheeler2004}, the model
principal components SMM1 and SMM2 are computed by projecting
the filtered data onto the two leading EOF modes and
standardizing the output such that a value of unity represents
an anomaly of $1$ standard deviation from the mean.

Strong MJO activity is characterized by an index amplitude
greater than or equal to 1. Precisely, following
\cite{Stachnik2015}, MJO events or episodes are defined from
the (SMM1,SMM2) time series as periods during which the
following conditions are met:
\begin{itemize}[leftmargin=30pt]
    \item The amplitude of SMM is greater than 1: $\sqrt{SMM1^2 + SMM2^2}\ge 1$.
    \item The propagation of the event is almost continually counterclockwise in the (SMM1, SMM2) space, corresponding to an almost continually eastward propagation (a westward propagation is limited to at most a single phase of the MJO cycle).
    \item The event propagates through at least $4$ phases {of the MJO cycle}.
\end{itemize}

\section{Numerical solution}
\label{sec:skel_model_num_sol}

\begin{sloppypar}
    In this section, we present the main features of the numerical solution to the stochastic MJO skeleton model \eqref{eq:skeleton-model-numerics} with (dimensionless) parameters  $\tilde{Q}=0.9$, $\Gamma=1.0$, $\bar{H}=0.22$, \blue{as in \cite{Ogrosky2015}}, and observation-based time-dependent sources of cooling and moistening $S^\theta$ and $S^q$. \blue{The spatio-temporal resolution of the model is chosen as in \cite{Thual2014}, with the spatial step $\Delta x \approx 625$ km (that is $40000/64$, where $40000$ km approximates the circumference of the Earth at the Equator) and the temporal step $\Delta t \approx 1.7$ h. Stochasticity is implemented by applying an independent replica of the stochastic process described in Sect. \ref{subsec:stoch-mod} to each spatial point $x$.}  Our Julia implementation of the model is available
from \cite{DataCode}. To make sure that the solutions are
presented for a statistically equilibrated regime, we run
simulations for 215 years with forcing corresponding to the
43-year period 1979-2021 repeated 5 times ($5 \times 43 =
215$). We then keep only the last 43 years which we consider
representative of the 1979-2021 period.
\end{sloppypar}

\subsection{Hovm\"oller diagrams of the model variables}

\begin{figure}[ht!]
\centering
\includegraphics[width = 0.75\textwidth]{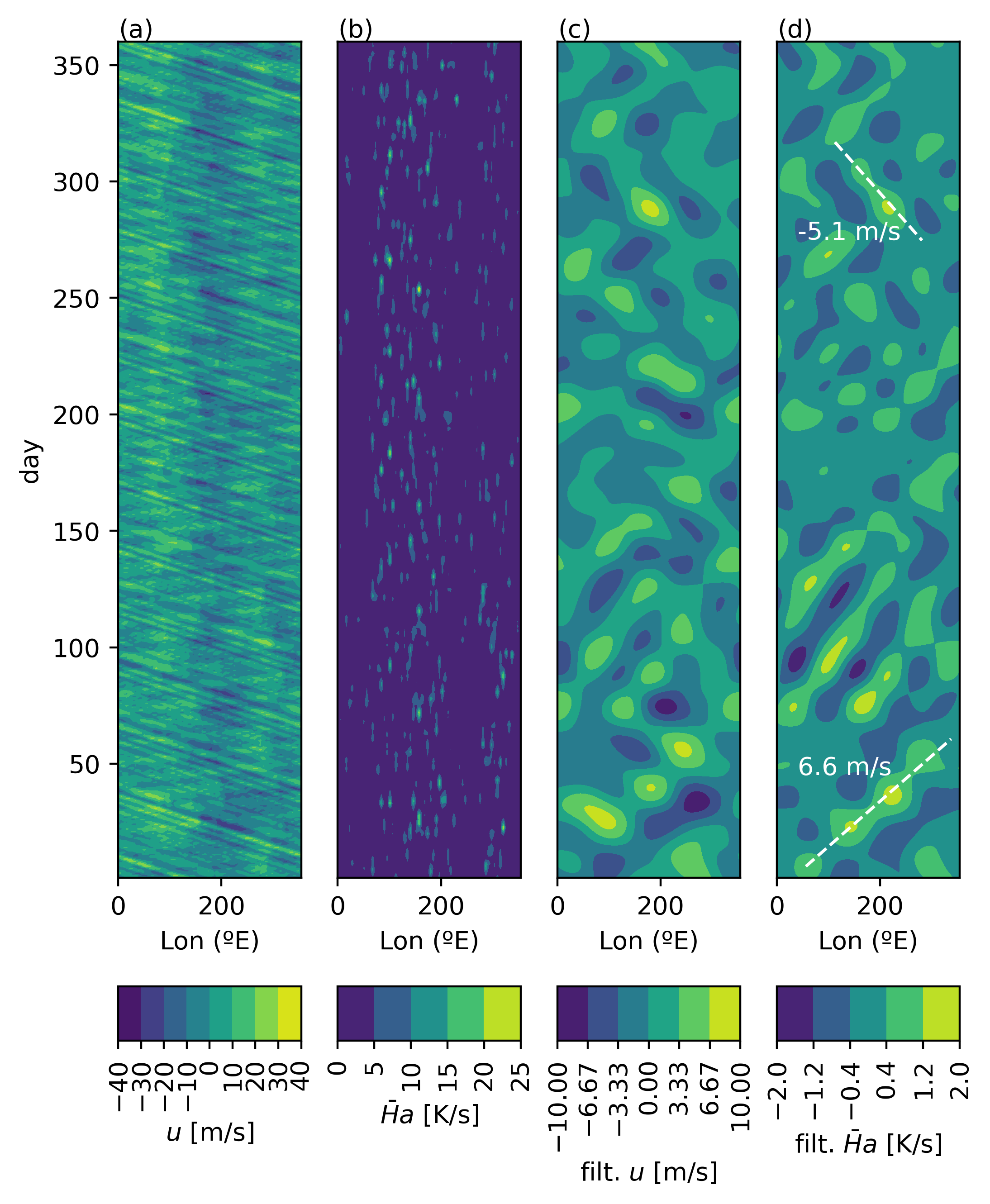}
\caption{Hovm\"oller diagrams of the skeleton model lower tropospheric wind $u(x,y=0,t)$ and envelope of convective activity $\bar{H}a(x,y=0,t)$ at the equator. (a,b) raw data, (c,d) daily anomalies from the long-term mean, filtered in time and space as described in the text. One westward moving signal (likely of a moist Rossby wave) and one eastward moving signal (likely MJO activity) are marked in white in panel (d).}
\label{fig:Hovmoller_skeleton_simulation}
\end{figure}

The evolution of the skeleton model equatorial profiles for the
lower tropospheric wind $u(x,t)$ and envelope of convective
activity $\bar{H}a(x,t)$, {computed from Eq. \eqref{eq:uvtheta}
with $y=0$}, are shown in Figure
\ref{fig:Hovmoller_skeleton_simulation}. The time axis
represents one year of simulation with forcing profiles
representative of the year 2005. Figure
\ref{fig:Hovmoller_skeleton_simulation}(a-b) represents the raw
output data. Figure
\ref{fig:Hovmoller_skeleton_simulation}(c-d) shows the data
after filtering in time and space as to isolate planetary-scale
intraseasonal variations. Precisely, the daily anomalies from
the long-term mean have been filtered in time using a $20-100$
days Lanczos filter and smoothed in space by retaining only
modes with Fourier wave number $k\le4$. While the non-filtered
plots [Fig. \ref{fig:Hovmoller_skeleton_simulation}(a,b)]
highlight the small scale propagating waves, the filtering
[Fig. \ref{fig:Hovmoller_skeleton_simulation}(c,d)] allows to
capture larger-scale features.

{Westward propagating modes are well visible in panel (a), as
well as in panel (c) and (d), e.g. around days 260 to 300. With
periods ranging from around 25 to 90 days, these modes could be
related to equatorial Rossby waves (see also Section
\ref{subsec:power_spec}). On the other hand, large-scale
eastward propagating waves are visible in panel (c) and (d),
especially towards the beginning of the period, from day 0 to
150. These intraseasonal large-scale waves are likely
associated with the MJO as we will see in Section
\ref{subsec:power_spec}.} Two waves, one propagating eastward
and one propagating westward, have been marked in Figure
\ref{fig:Hovmoller_skeleton_simulation}(d) with their
respective phase speed. Finally, we observe that in the
convective activity plots [Fig.
\ref{fig:Hovmoller_skeleton_simulation}(b,d)], a higher
activity can be seen in the region from $60^\circ$E to
$200^\circ$E which corresponds to a region between the Indian
Ocean and the western Pacific, which is the region where the
MJO signal is usually the strongest.

\subsection{Power spectrum of the envelope of convective activity}
\label{subsec:power_spec}

\begin{figure}[ht!]
\centering
\includegraphics[width = 0.5\textwidth]{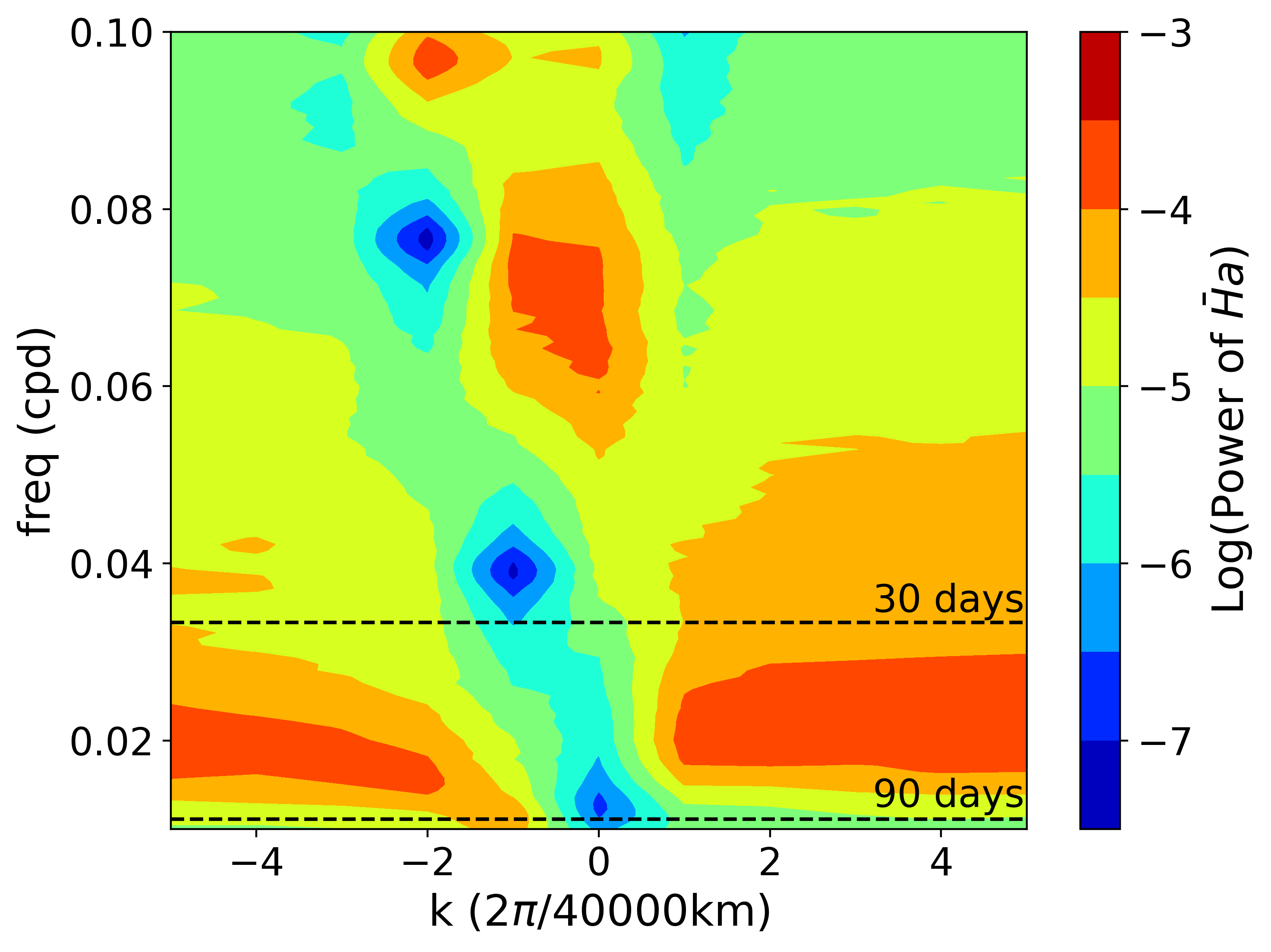}
\caption{Zonal wavenumber - frequency power spectrum of the simulated envelope of convective activity $\bar{H}a$ (in base 10 - logarithm). The dashed lines mark the 90 and 30 days periods.}
\label{fig:spectrum_MJO_model}
\end{figure}

Figure \ref{fig:spectrum_MJO_model} shows the zonal wavenumber
- frequency power spectrum of the simulated (unfiltered)
envelope of convective activity $\bar{H}a$. The zonal
wavenumbers are expressed as multiples of $2\pi/40000$ km
and frequencies are in cycles per day (cpd). The dashed lines
indicate the 90 and 30 days periods. The MJO appears as an
horizontally elongated high power structure in the zonal
wavenumber spectrum with $ 1 \leq k \leq 5$, that is as a
planetary-scale wave, with intraseasonal frequencies $1/90 \leq
\omega \leq 1/30 \textnormal{ cpd}$ and eastward propagation
($\omega/k > 0$). This structure has a dispersion relation
$d\omega/dk \approx 0$ which is a typical characteristic of the
MJO and is known to be reproduced well by the skeleton model
\citep{Majda2009b,Majda2011,Thual2014}. The mean phase speed of
the waves associated with this structure (calculated as the
mean of $\omega/k$ for the points with $\omega\in [1/90, 1/30]$
cpd, $k \in [1, 5]$ and log-power greater than $-4.0$) is
$\approx 5$ ms$^{-1}$. {This MJO signal is visible in Figure
\ref{fig:Hovmoller_skeleton_simulation}(d) between days 1 and
150.} In addition to the MJO signal there is also a high power
structure at intraseasonal time scales with westward
propagation. {Previous studies have shown that these modes
share some, although incomplete, features with convectively
coupled equatorial Rossby waves and they have been referred to
as moist Rossby modes \citep{Majda2011,Thual2014}. An example
of such westward wave is visible in the filtered convective
activity in Figure \ref{fig:Hovmoller_skeleton_simulation}(d)
between days 260 and 300.} At higher frequencies, $\omega >
0.06$ corresponding to periods shorter than 16 days, the high
power peaks might be associated with dry Kelvin and dry Rossby
modes \citep{Majda2011,Thual2014}. Overall the spectrum of
$\bar{H}a$ agrees well with previous studies of the skeleton
model \citep{Thual2014,Ogrosky2015}. {While several equatorial
modes and especially the MJO are well represented, many modes
are not reproduced by the model due to its minimal design, for
instance convectively coupled Kelvin waves
\citep{Kiladis2009}.}

\subsection{Climatology and variance of the envelope of convective activity}
\begin{figure}[ht!]
\centering
\includegraphics[width = \textwidth]{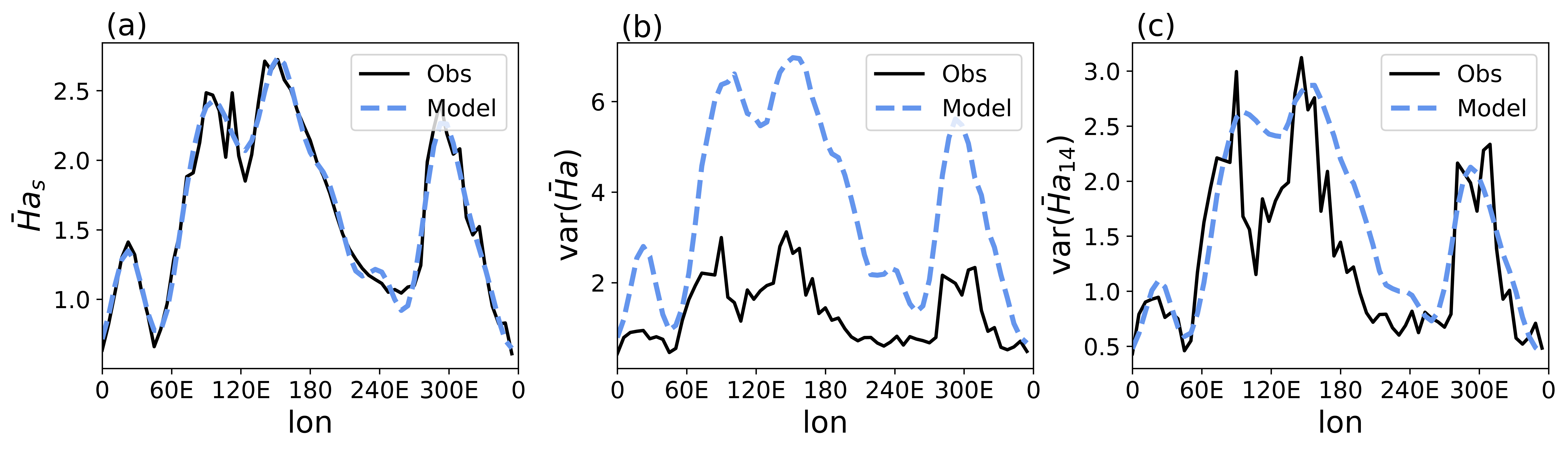}
\caption{(a) Long-term averages of observed and modeled $\bar{H}a$ (envelope of convective activity). (b) Variance of modeled and observed $\bar{H}a$ when all spatial modes from the model output are kept. (c) Variance of modeled and observed $\bar{H}a$ when only the first 14 spatial modes from the model output are kept. Note the different scales in panels (b) and (c).}
\label{fig:climatology_and_var_conv_act}
\end{figure}

The long-term means of $\bar{H}a$ estimated from daily observations and simulated in the skeleton model are shown in Figure \ref{fig:climatology_and_var_conv_act}(a). Qualitatively, they agree well. However, the variance of $\bar{H}a$ is overestimated in the model as can be seen in Figure \ref{fig:climatology_and_var_conv_act}(b). Nonetheless, if we consider only the first 14 spatial modes, the variance is well reproduced by the model (Fig. \blue{\ref{fig:climatology_and_var_conv_act}(c)}). This might be explained by the fact that the equations of the skeleton model are long-wave scaled, as the model is only concerned with planetary scales, and hence it might not be well suited to represent modes with higher wavenumbers \blue{(which are nevertheless continuously excited by the stochastic dynamics of the convective activity). The appropriate number of long-wavelength modes to be retained in the model's output in order to achieve agreement with observations has been determined empirically, and its justification would need a more detailed modeling approach in which smaller scales would be consistently included.}

\section{Identification of MJO events with the SMM index}
\label{sec:ID_MJO} The results presented above suggest the
presence of intermittent MJO wave structures propagating
eastward. In order to objectively identify these structures, we
use the skeleton multivariate MJO (SMM) index presented in
Section \ref{sec:rmm_index}.

\begin{figure}[ht!]
    \centering
    \includegraphics[scale = 0.55]{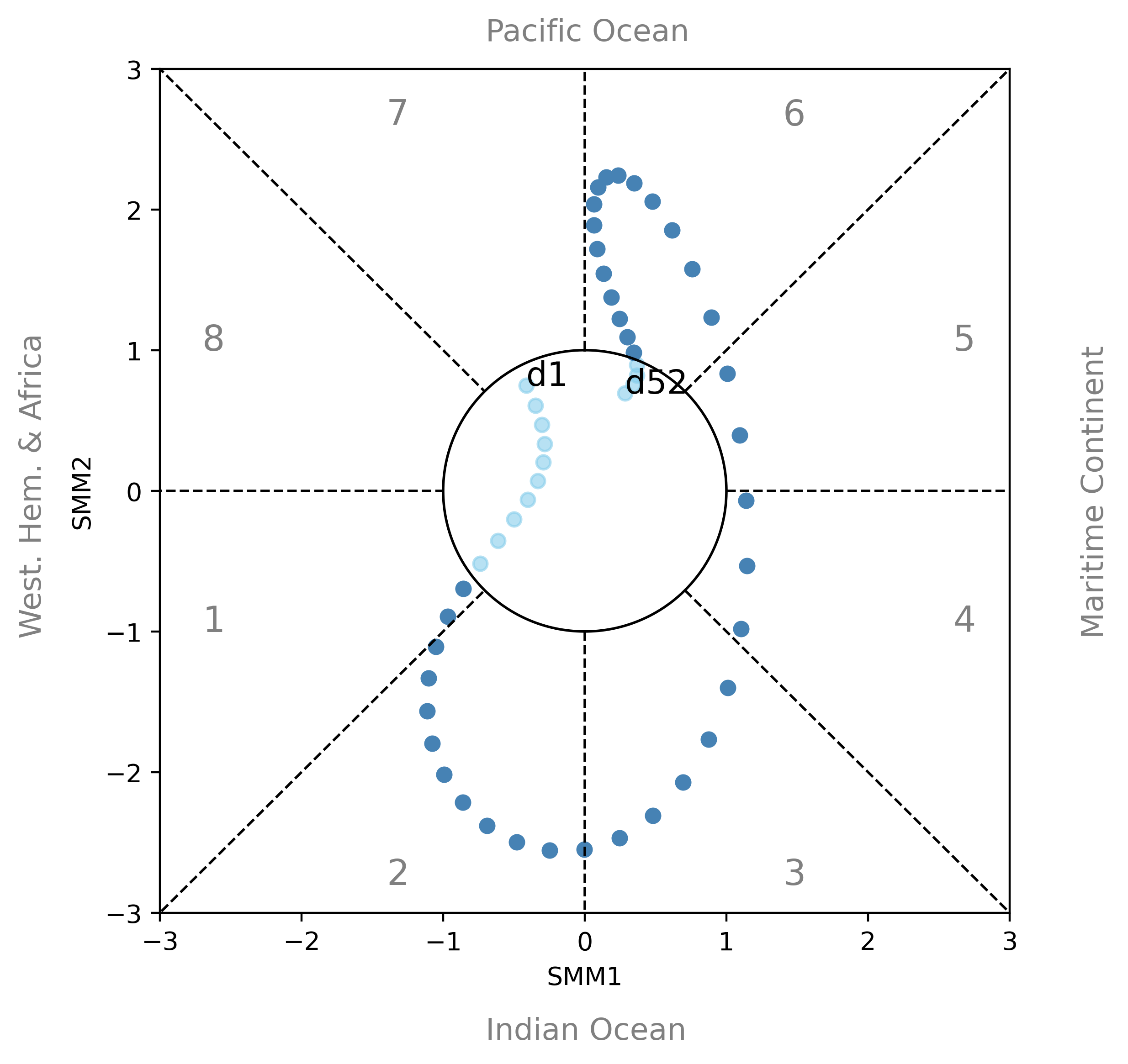}
    \caption{{Phase-space diagram of the model SMM values for a 52-day period from a simulation forced with observation-based functions. The dots correspond to daily values of (SMM1, SMM2). The first day of the series is labelled `d1', the last is labeled `d52'. The circle in the center of the plot has unit radius, indicating the threshold at which the amplitude of SMM exceeds $1$. Points in dark blue indicate an MJO event as defined by the criteria in Section \ref{sec:rmm_index}.}}
    \label{fig:skeleton_RMM_phase_space}
    \end{figure}

An example of model SMM values over a 52-day period are shown
in (SMM1, SMM2) phase space in Figure
\ref{fig:skeleton_RMM_phase_space}. The first and last point of
the series are annotated. The dark blue points satisfy the MJO
event's criteria given in Section \ref{sec:rmm_index}. Overall
the MJO propagation is relatively smooth. As explained in
\cite{Stachnik2015}, this is partly due to the filtering of
high frequencies in the model SMM index computation. This
filtering eliminates some of the day to day variability and
noise that are not removed in the computation of
observation-based RMM index from \cite{Wheeler2004}.

\section{MJO characteristics in the skeleton model and in observations}
\label{sec:obs_sim}

In the following, we study different characteristics of the MJO events in simulations and observations. The aim is to assess the ability of the MJO skeleton model to statistically reproduce the characteristics of observed MJO events. Here we list the chosen characteristics.

\begin{itemize}
     \item The \emph{seasonal variation} in the occurrence of MJO events is obtained by recording the number of MJO events occurring (that is starting, continuing or ending) during each month of the year, where events are defined according to the criteria listed in Section \ref{sec:rmm_index}.
     \item The \emph{duration} of an event is defined as the number of days from the first to the last day of the event.
     \item We measure the \emph{total angle} covered by an event tracked in the (SMM1, SMM2) phase space, that is the angle covered between the first and the last day of the event. This can roughly be assimilated to the ``distance'' covered by that event as it propagates along the equator.
     \item The \emph{maximum value of SMM amplitude}, where the amplitude is defined as $\sqrt{SMM1^2 + SMM2^2}$, is recorded for each event.
     \item Finally, the \emph{starting and ending phase of each event} {(1-8)} are recorded.
\end{itemize}

The model MJO events are identified using the SMM index and the
criteria described in Section \ref{sec:rmm_index}. We perform
$15$ independent simulation runs (in statistically equilibrated
regime) with forcing profiles representative of the period
1979-2021, leading to a total of $980$ modeled events. The
observed events are computed according to the same criteria
using the RMM index values which are freely available on the
website of the \citeauthor{RMM_data}
(\url{http://www.bom.gov.au/climate/mjo/}). We find $153$
observed events over the period 1979-2021. For these observed
events, the computation of the characteristics listed above is
made from the (RMM1,RMM2) values.

\subsection{MJO seasonal variations}

\begin{figure}[ht]
\centering
\includegraphics[width = 0.45\textwidth]{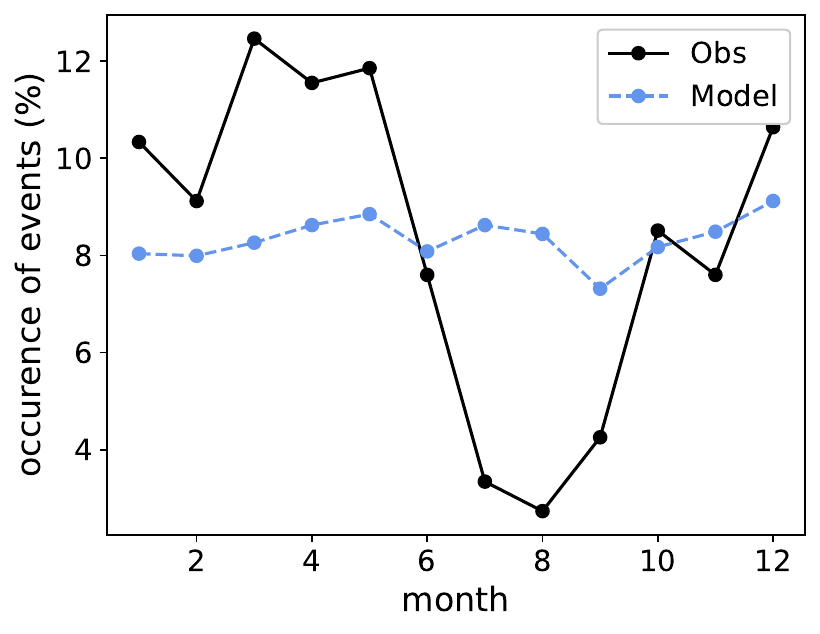}
\caption{Occurrence of MJO events as a function of the month of the year}
\label{fig:seasonality}
\end{figure}

We first look at the seasonal variation of MJO occurrences in
the model and in observations in Figure \ref{fig:seasonality}.
Observations show that MJO events are more frequent during
boreal winter and spring, from December through May. In the
simulations however no variation is detected. Recall that the
forcing profiles have been averaged with a 3-month window. As a
result the seasonal variations in heating and moistening are
smoother and \blue{of reduced amplitude, so that} the differences between different seasons might be
lost.

\subsection{MJO lifetime, extent in SMM/RMM phase space and maximum SMM/RMM amplitude}

\begin{figure}
\centering
\includegraphics[width = \textwidth]{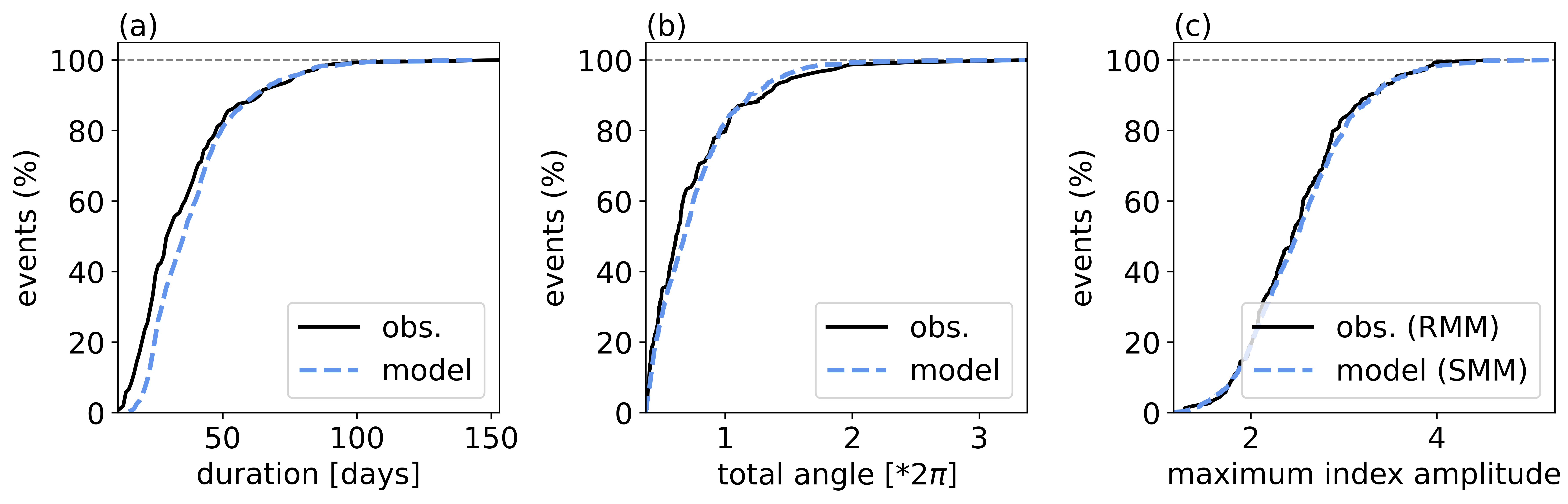}
\caption{Cumulative probability distribution of (a) the MJO events' durations, (b) total angle covered in  (RMM1, RMM2) / (SMM1, SMM2) phase space and (c) maximum RMM/SMM value for observed and simulated MJO events (981 simulated events and 153 observed events).}
\label{fig:MJO_characteristics_obs_sim}
\end{figure}

We now compare the duration of MJO events, total angle covered
in SMM/RMM phase space and maximum SMM/RMM amplitude in the
model outputs and in observations. The cumulative probability
distributions of these three characteristics are shown in
Figure \ref{fig:MJO_characteristics_obs_sim}. As above, the
statistics are based on $153$ observed events over the period
1979-2021 and $980$ modeled events from $15$ independent
simulations, run with forcing profiles representative of the
same period. The duration of events in Figure
\ref{fig:MJO_characteristics_obs_sim}(a), total angle
(distances) covered by simulated events in Figure
\ref{fig:MJO_characteristics_obs_sim}(b) and the maximum
amplitude of SMM/RMM in Figure
\ref{fig:MJO_characteristics_obs_sim}(c) compare well with
observations. The average MJO event lifetime is $39.6 \pm 0.8$
days for the simulation and $36.1$ days for observations. This
small difference is likely explained by the absence of certain
sources of MJO termination in the minimalistic skeleton model
leading to slightly longer events (although the longest event
overall, of 153 days, occurs in observational data). \blue{For comparison, the average event lifetime when time-independent forcings are used (i.e. when the model is forced with the long-term averages of the computed $S^q$ and $S^{\theta}$) is $40.9 \pm 0.7$ days, indicating a (small) improvement by using the more realistic time-dependent forcings.}  The mean angle is $(0.75 \pm 0.01)\cdot 2 \pi$ for simulations and $0.75\cdot2\pi$ for observations. \blue{This is an improvement with respect to the time-independent-forcing result of $(0.78 \pm 0.01) \cdot2\pi$}. The mean of the maxima of SMM amplitude is $2.53 \pm 0.02$ for the simulations \blue{(both with time-dependent and time-independent forcings)} and the mean of the maxima of RMM amplitude is $2.50$ for observations.

For each of the characteristics, a two-sample Kolmogorov-Smirnov test was conducted to compare the empirical cumulative distribution functions obtained from observations and from the \blue{time-dependent-forcing} model. \teal{The test's null hypothesis is that observed and modeled samples come from the same underlying distribution. For the duration, the test yields a $p$ value of $0.0006$, leading to the rejection of the null hypothesis and suggesting that the distributions might differ despite apparent similarities. For the total angle covered in the SMM/RMM phase space, $p=0.0428$, which also leads to the rejection of the null hypothesis at the $5\%$ significance level, although differences are less pronounced. For the maxima of SMM/RMM amplitude values, the $p$ value is $0.8116$, indicating that the model and observations produce statistically indistinguishable distributions.}

\subsection{MJO starting and ending phases}

\begin{figure}
\centering
\includegraphics[width = 0.85\textwidth]{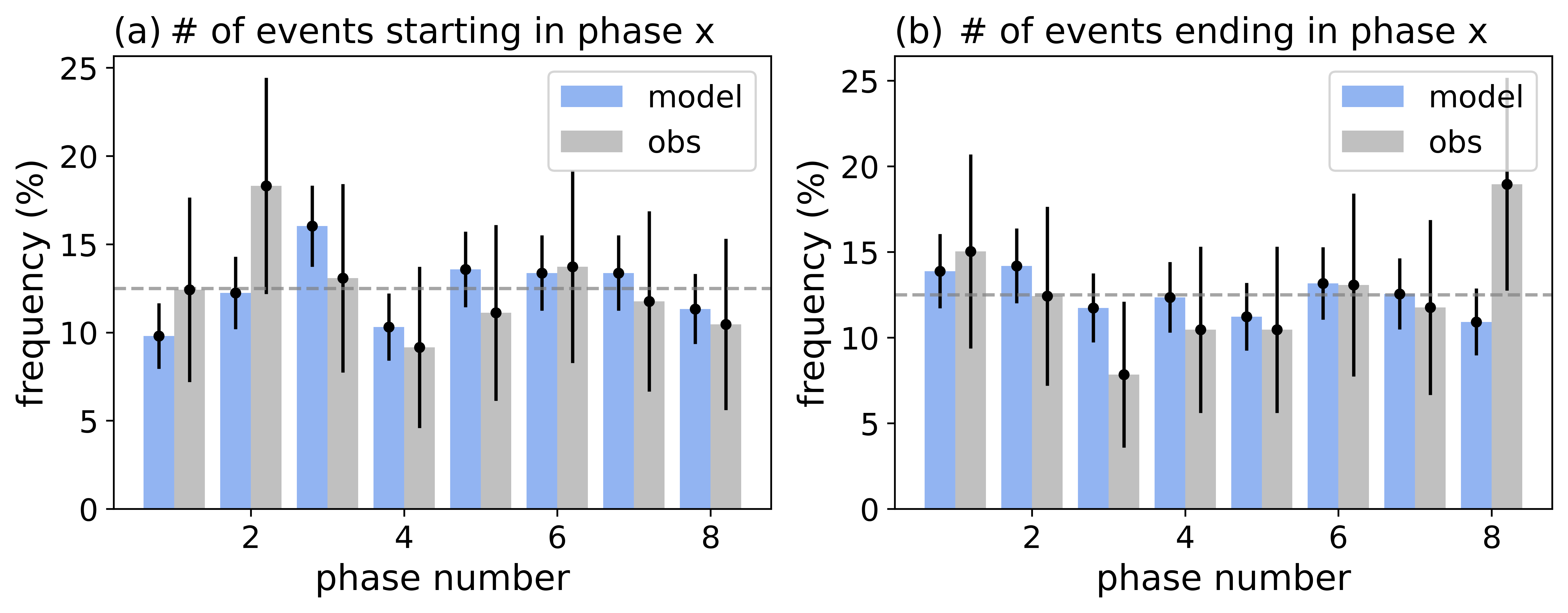}
\caption{Histograms of the recorded starting \emph{(a)} and ending phases \emph{(b)} of MJO events in the model and in observations. The dashed line indicates the equal likelihood of the ending location of an event being recorded in any of the 8 bins. The error bars indicate the $98\%$ uncertainty estimate, calculated from a binomial proportion confidence interval.}
\label{fig:MJO_starting_ending_phases}
\end{figure}

Figure \ref{fig:MJO_starting_ending_phases} shows the
distribution of initial and final phases of MJO events. The
error bar for a given bin is calculated using a binomial
proportion confidence interval dependent on the pass and fail
rate of recorded locations being assigned to that particular
bin. The dashed line indicates the equal likelihood of the
ending location of an event being recorded in any of the 8
bins. Almost all error bars overlap this line, both in the
simulations and observations, indicating that these graphs do
not allow for statistically significant conclusions to be
drawn. We note nonetheless that the distribution of starting
phases might have some similarities. Two local maxima are
observed around phases 2/3 and 6/7 for the starting location
(although a high peak is also shown in phase 5 for the model
which is not present in observations). For the ending phases,
in observations, most of the events end in phase 8, whereas in
the model the peak in phase 8 is relatively small. \bigbreak

\blue{To conclude Section \ref{sec:obs_sim}, the time-dependent
stochastic skeleton model captures \teal{reasonably} well statistical features
of the MJO events such as the distributions of durations, total
angles covered in SMM/RMM phase space and the maxima of SMM/RMM
values, despite \teal{some} disparities. However, it does not
reproduce well seasonal variations in the MJO occurrences.
Neither does it seem to be able to reproduce spatial
differences in the MJO properties such as its starting and
ending phase, although more investigation would be needed to
establish the presence of statistically significant
differences.} In the next section, we assess the ability of the
model to produce differences in the distributions of durations,
total angles and SMM maxima under El Ni\~no, La Ni\~na and
neutral ENSO conditions. Note that despite the model not being
able to reproduce seasonal variations, MJO events might still
be influenced by ENSO variations since they occur on much
longer time scales (typically of 2 to 7 years, although ENSO
itself also varies seasonally in intensity).

\section{Modulation of the MJO by ENSO}
\label{sec:ENSO_mjo_influence}

Equipped with our implementation of the stochastic skeleton model suitable
to incorporate time-dependent observation-based forcings, we
now study changes in the statistics of selected MJO
characteristics under the different phases of ENSO (El Ni\~no,
La Ni\~na and neutral phase) in observations and in the model.

To identify the ENSO phase during the period 1979-2021, we use
the NOAA Oceanic Ni\~no Index (ONI), based on SST anomalies in
the Ni\~no 3.4 region from $170^\circ-120^\circ$W and
$5^\circ$S-$5^\circ$N. The index is computed by averaging the
monthly values of Ni\~no 3.4 SST anomalies with a running
three-month window. An El Ni\~no event is declared when the
index is greater than or equal to $0.5^\circ$C for at least 5
consecutive values, \emph{i.e.} 5 consecutive overlapping
three-month seasons. A La Ni\~na event is declared when the
index is less than or equal to $-0.5^\circ$C for at least 5
consecutive values. The Ni\~no 3.4 values are available on the
website of \citeauthor{NINO_34_data}
(\url{https://origin.cpc.ncep.noaa.gov/products/analysis_monitoring/ensostuff/ONI_change.shtml}).

Identifying MJO events which occurred during each of ENSO phases is straightforward for observational data, based on the
event's date. Precisely, the whole length of the MJO event
is considered. An event is said to have occurred during El
Ni\~no/La Ni\~na/neutral ENSO, if more than half of the event
has occurred during that specific phase. For the model, as
explained in Section \ref{sec:skel_model_num_sol}, simulations
are run with forcing corresponding to the period 1979-2021,
resulting in time-stamped outputs which are considered
representative of this period. We can then also identify simulated
MJO events occurring during El Ni\~no, La Ni\~na and neutral
ENSO phases based on their dates. In addition, whereas in
observations the number of MJO events is constrained to a
single $43$-years record, limiting the significance of
statistical studies, we realize $15$ independent runs of the
model, obtaining much larger samples and more robust
statistical results. The total number of MJO events in
observations and in the model runs during each phase of ENSO is
reported in Table \ref{table:num_MJO_events}.

\begin{figure}[!ht]
    \centering
    \includegraphics[width = 0.65\textwidth]{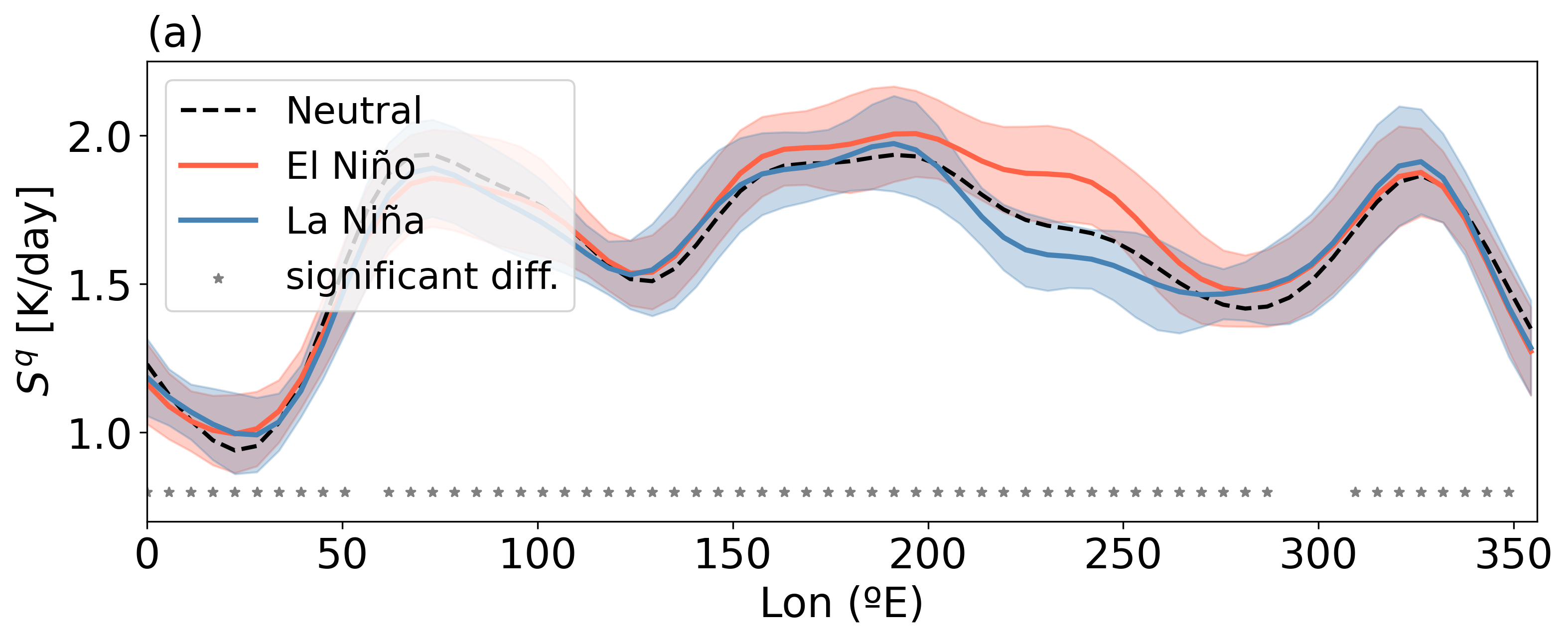}\\
    \includegraphics[width = 0.65\textwidth]{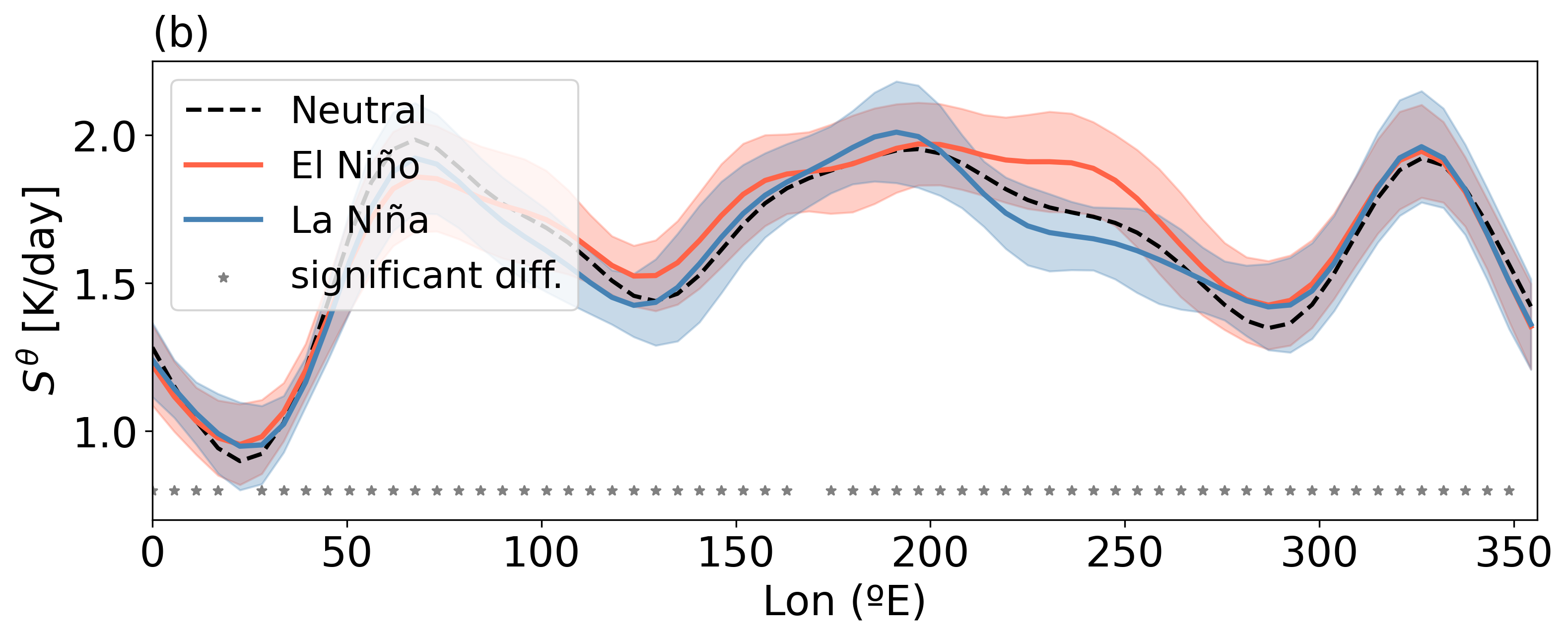}
    \caption{Mean profiles of (a) latent heating ($S^q$) and (b) radiative cooling
    ($S^\theta$), during El Ni\~{n}o, La Ni\~{n}a and neutral ENSO conditions from 1979 to 2021.
    The shaded areas correspond to 1 standard deviation around the means. Stars indicate locations
    where forcing profiles differ significantly (at the $5\%$ significance level) during
    El Ni\~{n}o and La Ni\~{n}a.}
    \label{fig:forcing-profiles-ENSO}
\end{figure}

\teal{To assess the ability of the stochastic skeleton model to
reproduce the observed modulation of MJO activity and
characteristics under different ENSO phases, it is essential
that the forcing functions $S^q$ and $S^{\theta}$ adequately
capture ENSO variability. Figure
\ref{fig:forcing-profiles-ENSO} shows that this is indeed the
case, by presenting the mean $S^q$ and $S^{\theta}$ profiles
during El Ni\~{n}o, La Ni\~{n}a and neutral ENSO conditions.
The largest differences are effectively observed in the eastern
equatorial Pacific. Furthermore, we used the Kolmogorov-Smirnov
test to assess, at each spatial point, whether the
distributions of $S^q$ ($S^{\theta}$) values differ
significantly during El Ni\~{n}o and La Ni\~{n}a. We found
significant differences at nearly all points at the $5\%$
significance level (indicated by stars in the figure). }

\begin{table}[!ht]
\renewcommand{\arraystretch}{1.1}
\setlength\tabcolsep{0pt}
\centering
\begin{tabular*}{\textwidth}{@{\extracolsep{\fill}}|P{0.32} |P{0.16}| P{0.16}| P{0.16}| P{0.16}| }
\hline
ENSO phase & El Ni\~no & La Ni\~na & neutral & Total\\
\hline
\hline
Observed MJO events  & $42$  &  $36$  &  $75$ &  $153$ \\
\hline
Simulated MJO events & $205$ &  $228$  & $547$ &  $980$\\
\hline

\end{tabular*}
\caption{Number of MJO events in observations and in the model, during each phase of ENSO. }
\label{table:num_MJO_events}
\end{table}


\subsection{MJO activity across El Ni\~no, La Ni\~na and neutral ENSO}

We first report the occurrence of MJO events across ENSO phases
for observations and simulations. Table
\ref{table:num_days_MJO_events} shows the percentage of MJO
active days during El Ni\~no, La Ni\~na and neutral ENSO for
the period 1979-2021 (in observations and simulations). Note
that, for the simulations, the values correspond to a mean over
the 15 independent runs, and the standard error of the mean is
indicated. In addition, the total number of days (irrespective
of MJO activity) belonging to each of the three phases is
indicated in the last row.  We observe that the proportion of
MJO days approximately follows the proportion of the total
number of days belonging to each phase of ENSO both for
observations and simulations, with the majority of events
occurring during the neutral phase of ENSO, and the minimum
during El Ni\~no. When comparing percentages in the first row
with those in the last row, we see that, in observations, El
Ni\~no and La Ni\~na conditions seem to slightly favor the
occurrence of MJO events (with respect to what would be
expected from the proportion of these ENSO phases). On the
other hand, when comparing the second row to the last, we see
that, in the simulations, the opposite is observed: the neutral
ENSO conditions seem to slightly favor the occurrence of MJO
events. This small discrepancy could be attributed to the
design of the model, i.e. the omission of certain mechanisms
and interactions with other climate phenomena, {though the
differences are minor and may simply reflect limited sample
sizes.} \vspace{0.5cm}

\begin{table}[!ht]
\setlength\tabcolsep{0pt}
\centering
\begin{tabular*}{\textwidth}{@{\extracolsep{\fill}}|P{0.28} |P{0.239}| P{0.239}| P{0.239}| }
\hline
ENSO phase & El Ni\~no & La Ni\~na & neutral  \\
\hline
\hline
Observed active MJO days & $24.8\%$ & $26.3\%$ & $48.9\%$  \\
\hline
Simulated active MJO events & $21.7\pm 1.5\%$ & $23.7\pm 1.16\%$ & $54.6\pm 1.9\%$  \\
\hline
Total number of days per phase & $3497$ ($22.8\%$)  &  $3924$ ($25.6\%$)  &  $7920$ ($51.6\%$)  \\
\hline
\end{tabular*}

\caption{Comparison of observed and simulated percentages of
active MJO days across ENSO phases -- percentages from the
total of active MJO days, where ``active MJO days'' are the
days during MJO events as defined in Section
\ref{sec:rmm_index}. The last line represents the total number
of days (active and inactive MJO) in each phase of ENSO over
the period 1979-2021.} \label{table:num_days_MJO_events}
\end{table}

\subsection{ENSO modulation of MJO characteristics}

\subsubsection{Observations}

\begin{figure}[!ht]
\includegraphics[width = \textwidth]{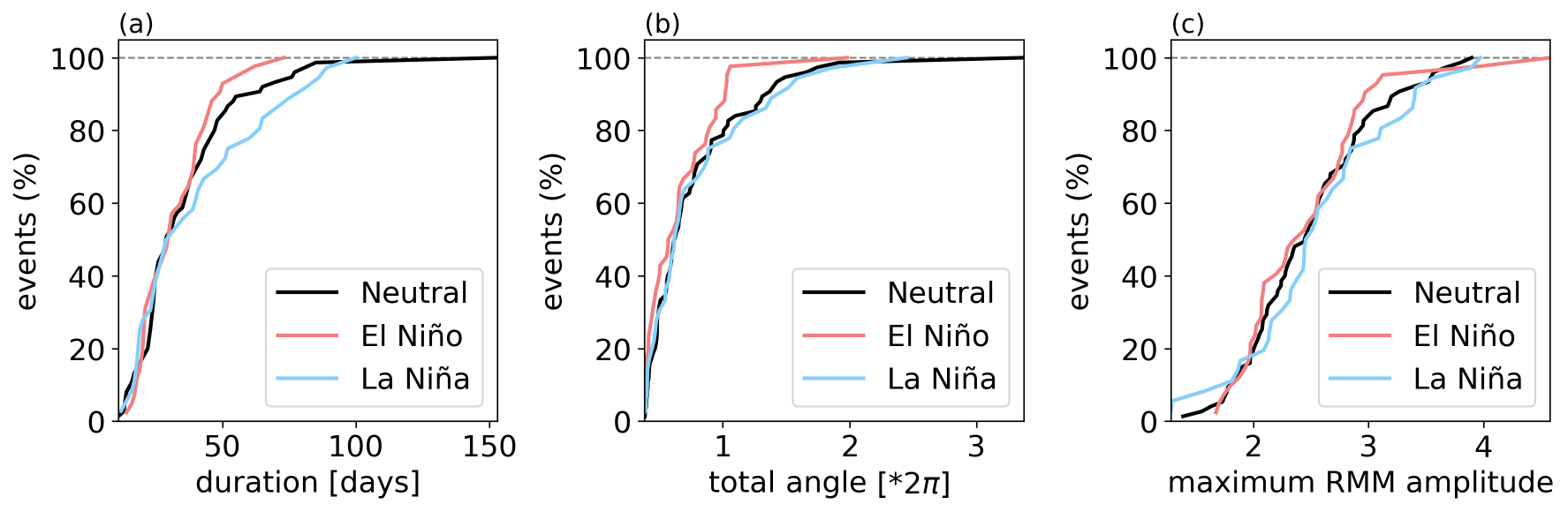}
\caption{Cumulative probability distribution of (a) the MJO events' durations, (b) total angle covered in (RMM1, RMM2) phase space and (c) maximum RMM value for \emph{observed} events occurring during El Ni\~no, La Ni\~na and neutral periods between 1979 and 2021. The number of events in each sample is reported in Table \ref{table:num_MJO_events}.}
\label{fig:MJO_characteristics_obs}
\end{figure}

Figure \ref{fig:MJO_characteristics_obs} shows the cumulative
probability distribution of the MJO events' durations, total
angle covered in (RMM1, RMM2) phase space and maximum RMM value
for observed events occurring during El Ni\~no, La Ni\~na and
neutral periods. The distributions show several differences.
During El Ni\~no, observed MJO events appear to have a shorter
duration compared to those occurring during La Ni\~na and
neutral ENSO [Fig. \ref{fig:MJO_characteristics_obs}(a)]. In
fact, the mean duration of MJO events during El Ni\~no is 33
days, 40 days for La Ni\~na and 36 days for the neutral phase
of ENSO. Further, the maximum duration is 73 days for MJO
events during El Ni\~no, 100 days during La Ni\~na and 153 days
during the neutral phase of ENSO. Similarly, during El Ni\~no,
the total angle covered in RMM phase space by MJO event seems
to be shorter [Fig. \ref{fig:MJO_characteristics_obs}(b)],
suggesting that events propagate over shorter distances. In
fact, the mean angle covered in RMM phase space by MJO events
occurring during El Ni\~no is $0.7\cdot 2 \pi$, while it is
$0.8\cdot 2 \pi$ for la Ni\~na and the neutral phase of ENSO.
The maximum angle covered in RMM phase space is $2.0\cdot 2
\pi$ for events occurring during El Ni\~no (meaning that they
have propagated twice around the entire globe), $2.4\cdot 2
\pi$ for events occurring during La Ni\~na and $3.4\cdot 2 \pi$
for events occurring during the neutral phase of ENSO. Note
that previous studies have reported that during El Ni\~no
periods, MJO events tend to propagate further eastward
\citep{Kessler2001,Tam2005,Pohl2007}. This does not contradict
the present results since the methodologies and definition of
MJO events vary between these studies. In addition, we consider
here the total angle in RMM phase space and not the final
location of MJO events. Finally, for the maxima of RMM
amplitude [Fig. \ref{fig:MJO_characteristics_obs}(c)], the
differences in the cumulative distributions are slightly more
intricate than for the other two characteristics. The mean is
$2.5$ for events during El Ni\~no, $2.6$ during the La Ni\~na
and $2.5$ during the neutral phase. The maximum values are
$4.6$ for El Ni\~no, $4.0$ for La Ni\~na, and $3.9$ for the
neutral phase of ENSO. The mean and maximum values of all
characteristics are summarized in Table
\ref{tab:MJO_comparison}.

\begin{figure}[!ht]
\includegraphics[width = \textwidth]{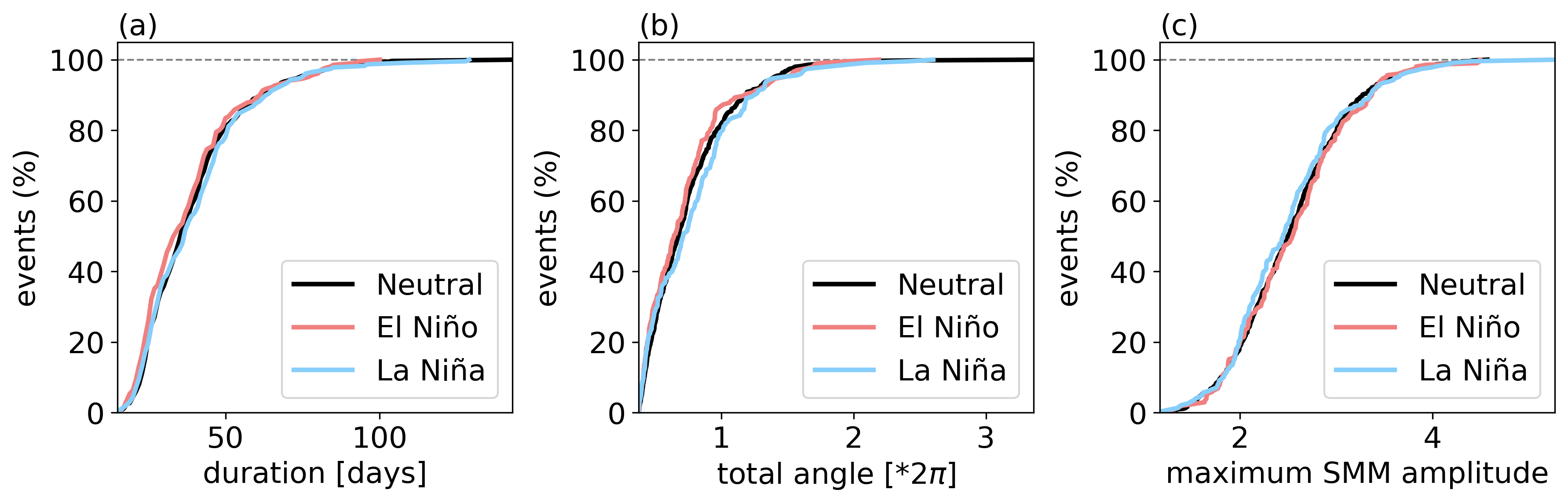}
\caption{Cumulative probability distribution of the MJO events' durations, total angle covered in (SMM1, SMM2) phase space and maximum SMM value for \emph{simulated} events occurring during El Ni\~no, La Ni\~na and neutral periods. The simulations were forced with data covering the period 1979-2021. The number of events in each sample is reported in Table \ref{table:num_MJO_events}.}
\label{fig:MJO_characteristics_sim}
\end{figure}

\subsubsection{Simulations}

Finally, we look at the characteristics of MJO events in the
model, during El Ni\~no, La Ni\~na and neutral conditions. The
results are shown in Figure \ref{fig:MJO_characteristics_sim}.
For all three MJO characteristics, the cumulative distributions
are very similar during El Ni\~no, La Ni\~na and neutral
conditions. We perform a Kolmogorov-Smirnov test to identify
whether statistically significant differences exist between the
El Ni\~no and La Ni\~na samples for each of the three
characteristics. For the duration of MJO events, the test
yields a $D$ value (representing the maximum distance between
the two curves) of 0.0865 and a $p$ value of 0.4. For the total
angles covered, the $D$ value is 0.1307 and the $p$ value 0.05.
For the maxima of SMM values, the $D$ value is 0.0858 and the
$p$ value 0.4. Hence, none of the three tests provides
sufficient evidence to reject the null hypothesis that the two
samples come from the same distribution at the $5\%$ significance level.

We might nonetheless compare the mean and maxima of the three
characteristics. To do so we compute these values in 15
independent simulation runs and determine the mean and standard
error of the resulting values. They are summarized in
Table~\ref{tab:MJO_comparison}. On average, we observe slightly
smaller mean and maximum durations of MJO events during El
Ni\~no compared to other phases of ENSO. We also observe
smaller mean and maximum values (on average) of the total angle
covered in SMM phase space during El Ni\~no. No specific
tendency is observed for the maxima of SMM values. In all
cases, the differences between these values remain small.

\begin{table}[ht]
    \setlength\tabcolsep{0pt}
    \centering
    \begin{tabular*}{\textwidth}{@{\extracolsep{\fill}}|P{0.22} |P{0.145}| P{0.21}| P{0.21}| P{0.21}| }
    \hline
    \textbf{Parameter} & \cellcolor{white}\textbf{Statistic} & \multicolumn{1}{>{\columncolor{white}}c|}{\textbf{El Ni\~no}} & \multicolumn{1}{>{\columncolor{white}}c|}{\textbf{La Ni\~na}} & \multicolumn{1}{>{\columncolor{white}}c|}{\textbf{neutral}} \\
    \hline
    \multirow{6}{=}{\centering Duration (days)} &  \small{\textit{Obs.}} &   &   &    \\
    &  \small{Mean} &  33 &  40 &  36  \\
    &  \small{Max.} &  73 &  100 &  153  \\
    \cline{2-5}
    & \small{\textit{Sim.}} &   &   &    \\
    &  \small{{Mean}} &  $38.3\pm1.2$ &  $40.7\pm1.9$ &  $39.6\pm0.8$ \\
    &  \small{{Max.}} &   $73.4\pm4.4$ &  $80.9\pm6.2$ &  $90.7\pm5.6$ \\
    \hline
    \multirow{6}{=}{\centering Angle in SMM Phase Space} &  \small{\textit{Obs.}} &   &   &    \\
    &  \small{{Mean}} &  $0.7\cdot 2 \pi$ &  $0.8\cdot 2 \pi$ &  $0.8\cdot 2 \pi$ \\
    &  \small{{Max.}} &  $2.0\cdot 2 \pi$ &  $2.4\cdot 2 \pi$ &  $3.4 \cdot 2 \pi$  \\
    \cline{2-5}
    & \small{\textit{Sim.}} &   &   &    \\
    &  \small{{Mean}} &   $(0.73\pm0.02)2\pi$ &  $(0.79\pm0.04)2\pi$ &  $(0.75\pm 0.01)2\pi$ \\
    &  \small{{Max.}} &   $(1.44\pm0.10)2\pi$ &  $(1.55\pm0.13)2\pi$ &  $(1.75\pm 0.15)2\pi$ \\
    \hline
    \multirow{6}{=}{\centering Maxima of SMM Amplitude} &  \small{\textit{Obs.}} &   &   &    \\
    &  \small{{Mean}} &  2.5 &  2.6 &  2.5 \\
    &  \small{{Max.}} &  4.6 &  4.0 &  3.9  \\
    \cline{2-5}
    & \small{\textit{Sim.}} &   &   &    \\
    &  \small{{Mean}} &  $2.57\pm0.04$ &  $2.51\pm0.05$ &  $2.53\pm 0.02$ \\
    &  \small{{Max.}} &   $3.71\pm0.11$ &  $3.82\pm0.14$ &  $3.96\pm 0.09$ \\
    \hline
    \end{tabular*}
    \caption{Comparison of the mean and maximum values of selected MJO characteristics between observations and simulations. For observations, the values are calculated as a mean over 15 independent simulation runs with the corresponding standard error.}
    \label{tab:MJO_comparison}
\end{table}

\subsubsection{Discussion}
Comparing observations and simulations, the main difference
lies in the fact that, in the simulations, the cumulative
distributions of MJO characteristics do not significantly
differ under different ENSO conditions (unlike in observations,
\teal{although limited samples sizes might affect observed
patterns}). \teal{Since the model's forcings $S^q$ and
$S^{\theta}$ are significantly different during El Ni\~{n}o and
La Ni\~{n}a conditions (Figure
\ref{fig:forcing-profiles-ENSO}), the lack of the model's
response to ENSO variability suggests a limitation in the
representation of key physical mechanisms governing the
MJO-ENSO interactions.}


However, some similarities between observations and simulations
are still seen when looking at the mean and maxima of MJO
characteristics. In both observations and the model, the mean
and maximum duration of MJO events are smaller for events
occurring during El Ni\~no and larger for events occurring
during La Ni\~na (although these differences are relatively
small in the model). Similarly, the mean and maximum of total
SMM/RMM angle are smaller during El Ni\~no than during La
Ni\~na. On the other hand, no such general tendency is seen for
the maximum of SMM/RMM amplitude.

\section{Conclusions}  
{We have implemented time-varying observation-based forcing
profiles in the MJO stochastic skeleton model. As in previous
works \citep{Majda2009b,Majda2011,Thual2014}, the model
captures several important features of the MJO including its
phase speed of around 5 m/s, its flat dispersion relation, its
horizontal quadrupole vortex structure (not shown here) and the
intermittent generation of MJO events. We saw that, when
considering planetary scales, the climatology and variance of
observed convective activity are in very good agreement. Using
an RMM-like index in the model we were able to objectively
identify MJO events and to study their characteristics. We find
agreements between observations and simulations for the
statistics of MJO event durations, total angles covered in
SMM/RMM phase space and maxima of SMM/RMM amplitudes. However,
we also showed that the model cannot reproduce well seasonal
variations in the MJO occurrences. \teal{This could be due to
the 3-month averaging of the forcing profiles, which was
introduced to ensure generation of MJO events in the model.
This forcing better captures the long-term trends, but may have
the side effect of blurring the differences between seasons. We
note that the number of MJO events generated in the stochastic
skeleton model forced with these profiles remains lower than in
observations. Future work should investigate the effect of
different lengths of time-averaging windows on the frequency of
MJO events in the model and on their statistical properties, to
find a right balance between the large-scale nature of the
model ingredients, and an appropriate inclusion of relevant
short-scale variability.}

The model also fails to accurately replicate spatial variations in the MJO
properties, such as its starting and ending phase. This
limitation may be attributed to the stochasticity in the model,
as pointed out in \cite{Stachnik2015}. The stochasticity likely
has the effect of dampening the geographic dependencies that
should arise from the application of zonally varying forcing
functions. Further investigation would be required to separate
the effects of stochasticity, zonally varying forcing and
nonlinearity on the spatial properties of simulated MJO events.
To investigate how temporal variability at longer time scales
affects the MJO, we evaluated differences in the statistics of
\blue{the three MJO characteristics mentioned above (duration,
angle and maximum amplitude)} under the different phases of
ENSO (El Ni\~no, La Ni\~na and neutral phase) in observations
and in the model. We found that while observations might
suggest some differences (for instance a tendency towards
shorter-lived MJO events during El Ni\~no), the model does not
identify statistically significant differences in duration,
total angle covered in SMM phase space and in the maximum SMM
value of simulated MJO events between the different ENSO
phases. }

{In conclusion, our results show that the model reproduces well
the planetary-scale variability of convective activity and
selected characteristics of MJO events, but it does not capture
the impact of ENSO phases on these characteristics. \blue{This
is due to limitations in the structure and ingredients of the
skeleton model.  More complex interactions such as changes in
mean state winds and ocean coupling \citep{Diaz2023,
Suematsu2022, Wei2019, Moser2024} are needed to capture the
interaction between ENSO and the MJO.} Further investigation
will be needed to determine whether the interannual variability
of the model forcing functions impacts other MJO
characteristics which have not been studied here, such as the
propagation speed of individual events or their longitudinal
extent. In order to compute these characteristics as precisely
as possible and to make objective comparisons between the model
and observations, one will need to implement a method able to
track the MJO (e.g. its convective center) both in the
simulations and reanalysis.

Further, while the skeleton model allows to gain a better
understanding of the fundamental physical mechanisms of the
MJO, more complexity will likely be needed to fully reproduce
the modulation of the MJO by ENSO. In particular, ENSO impacts
the extent of MJO penetration in the Pacific Ocean
\citep{Pohl2007,Tam2005,Kessler2001}. During El Ni\~no years,
MJO convective activities often extend eastward beyond the
dateline into the Pacific, while during La Ni\~na years, they
tend to remain west of the dateline. \cite{Pohl2007} also
reported that the duration of MJO events is longer during La
Ni\~na and shorter during El Ni\~no when events occur from
March to May and October to December. Hence, in order to be
able to study the effects of ENSO on the MJO, one condition on
any model should be that it captures the spatial variability of
the MJO as well as its seasonal variations. }


\codedataavailability{

The datasets used for computing the model's forcing profiles
(Section \ref{sec:forcing-profiles}) are publicly accessible.
The NCEP-NCAR reanalysis latent heat net flux data
\citep{Kalnay1996} can be freely downloaded from
\href{https://psl.noaa.gov/data/gridded/data.ncep.reanalysis.html}{NOAA
PSL}, while the NCEP Global Precipitation Climatology Project
(GPCP) data \citep{NCEPPrecip,Huffman2001} is available from
the
\href{https://www.ncei.noaa.gov/products/climate-data-records/precipitation-gpcp-monthly}{NOAA
NCEI}. The Python code used to compute these profiles is
accessible on
\cite{DataCode}. The Julia implementations of the stochastic
MJO skeleton model and the code used for the postprocessing and
analysis of its outputs are also available on \cite{DataCode},
as well as the model output data and Julia notebooks to
reproduce the figures in the present paper.


For observational indices: The RMM index data can be freely
downloaded from the website of the
\href{http://www.bom.gov.au/climate/mjo/}{Australian Bureau of
Meteorology}. The Ni\~no3.4 index can be freely accessed on the
website of
\href{https://origin.cpc.ncep.noaa.gov/products/analysis_monitoring/ensostuff/ONI_change.shtml}{NOAA
(National Centers for Environmental Information)}.

} 










\authorcontribution{N.E. performed the simulations, analyzed the data, created the
figures, and wrote the first manuscript draft.

N.E., R.V.D., E.H.-G., C.L. directed the study. All authors contributed with ideas,
interpretation of the results and manuscript revisions. } 

\competinginterests{At least one of the (co-)authors is a member of the editorial board of Nonlinear Processes in Geophysics.} 


\begin{acknowledgements}
N.E. would like to thank Nan Chen and Tabea Gleiter for their
help with understanding and implementing the stochastic
skeleton model. She is also grateful to H. Reed Ogrosky and
Samuel N. Stechmann for the helpful email exchanges.
    This project has received funding from the European Union's Horizon 2020 research and
    innovation programme under the Marie Skolodowska-Curie Grant Agreement No 813844,
    from the Agencia Estatal de Investigaci\'{o}n (MICIU/AEI/10.13039/501100011033) under
    the Mar\'{\i}a de Maeztu project CEX2021-001164-M, and
    from the Agencia Estatal de Investigaci\'{o}n (MICIU/AEI/10.13039/501100011033) and FEDER ``Una manera de hacer
    Europa" under Project LAMARCA No. PID2021-123352OB-C32. R.V.D. acknowledges financial
    support by the German Federal Ministry for Education and Research (BMBF) via
    the JPI Climate/JPI Oceans Project ROADMAP (Grant No. 01LP2002B).
\end{acknowledgements}


\bibliographystyle{copernicus}
\bibliography{bibliography}

\end{document}